\documentclass[journal]{IEEEtran}
\usepackage{amsmath,amsfonts}
\usepackage{algorithmic}
\usepackage{algorithm}
\usepackage{array}
\usepackage[caption=false,font=normalsize,labelfont=sf,textfont=sf]{subfig}
\usepackage{textcomp}
\usepackage{stfloats}
\usepackage{url}
\usepackage{verbatim}
\usepackage{graphicx}
\usepackage{cite}
\graphicspath{{Figures/}}
\usepackage{epstopdf}
\usepackage{gensymb}
\usepackage{mathtools}

\begin{document}

\title{3D terrain mapping and filtering from coarse resolution data cubes extracted from real-aperture 94 GHz radar}

\author{William D. Harcourt, David G. Macfarlane,~\IEEEmembership{Member,~IEEE,}, Duncan A. Robertson,~\IEEEmembership{Member,~IEEE}
        % <-this % stops a space
\thanks{This paper was produced by William D. Harcourt, David G. Macfarlane and Duncan A. Robertson based in the School of Physics and Astronomy at the University of St Andrews. William D. Harcourt is now based at the University of Aberdeen.}% <-this % stops a space
\thanks{Manuscript received May 1, 2023; revised TBC}}

\markboth{IEEE Transactions on Geoscience and Remote Sensing,~Vol.~00, No.~0, May~2023}%
{Shell \MakeLowercase{\textit{et al.}}: A Sample Article Using IEEEtran.cls for IEEE Journals}

%\IEEEpubid{0000--0000/00\$00.00~\copyright~2023 IEEE}

\maketitle

\begin{abstract}
Accurate, high-resolution 3D mapping of environmental terrain is critical in a range of disciplines. In this study, we develop a new technique, called the \textit{PCFilt-94} algorithm, to extract 3D point clouds from coarse resolution millimetre-wave radar data cubes and quantify their associated uncertainties. A technique to non-coherently average neighbouring waveforms surrounding each AVTIS2 range profile was developed in order to reduce speckle and was found to reduce point cloud uncertainty by 13\% at long range and 20\% at short range. Further, a Voronoi-based point cloud outlier removal algorithm was implemented which iteratively removes outliers in a point cloud until the process converges to the removal of 0 points. Taken together, the new processing methodology produces a stable point cloud, which means that: 1) it is repeatable even when using different point cloud extraction and filtering parameter values during pre-processing, and 2) is less sensitive to over-filtering through the point cloud processing workflow. Using an optimal number of Ground Control Points (GCPs) for georeferencing, which was determined to be 3 at close range ($<$1.5 km) and 5 at long range ($>$3 km), point cloud uncertainty was estimated to be approximately 1.5 m at 1.5 km to 3 m at 3 km and followed a Lorentzian distribution. These uncertainties are smaller than those reported for other close-range radar systems used for terrain mapping. The results of this study should be used as a benchmark for future application of millimetre-wave radar systems for 3D terrain mapping.
\end{abstract}


\section{Introduction} \label{Section_Introduction}
% INTRO TO 3D POINT CLOUDS
\IEEEPARstart{M}{apping} the changing shape and structure of Earth surface phenomena is vital for understanding environmental processes. Active sensors, such as radar or lidar, scan across a scene of interest at multiple azimuth $\left(\theta\right)$ and elevation $\left(\phi\right)$ angles, measuring the reflected signal intensity at different range $\left(R\right)$ bins. This leads to the construction of a 3D data cube of signal backscatter $\left(R,\theta,\phi\right)$. 3D point clouds, which represent the terrain geometry in $x$, $y$, and $z$, are extracted from these data cubes and are used to quantify Earth surface topography. They have been most widely derived from Terrestrial Laser Scanners (TLS) and Lidar \cite{Mallet2009,Smith2015,Okyay2019} where the data cubes are denser compare to those extracted from radar measurements due to the higher angular resolution of electro-optical techniques. Developing methodologies to extract terrain elevation from these 3D data cubes and their associated uncertainties is critical in determining the significance of spatial patterns that exist in such data sets, as well as the ability to detect changes when comparing multi-temporal point clouds. 

% MM-WAVE RADAR TERRAIN MAPPING
Millimetre-wave radar, which operates in the frequency range 30 to 300 GHz (wavelengths of 10 mm to 1 mm), can map terrain at high angular resolution for a given aperture size in conditions of reduced visibility using compact systems \cite{Currie1987}. These higher radar frequencies have seldom been used for terrain mapping due to its weaker performance at both long range and in adverse weather conditions compared to conventional lower frequency radar systems. However, the smaller physical size of millimetre-wave radar systems enables their deployment onto vehicles for short-range perception (i.e. up to a few km) and target detection when visibility is obscured. In this configuration, they have been used to generate real-time maps of quarries (both at the surface and underground) during dusty conditions in order to aid navigation and reduce the probability of collision with machinery and quarry cliffs \cite{Widzyk2006,Brooker2007}. Similarly, helicopter propellers create dust storms during landing, known as `brown-out', and millimetre-wave radars have enabled real-time surface mapping (synthetic vision) in Degraded Visual Environments (DVE) via helicopter-mountable radar systems \cite{Schmerwitz2011a,Schmerwitz2011b}. They can also be used to detect objects such as pylons to ensure safe landing \cite{Goshi2012} and for 3D mapping using Interferometric Synthetic Aperture Radar (InSAR) \cite{Magnard2014}. 

% AVTIS2 INTRODUCTION
Long-range 3D mapping of terrain up to 7 km has been demonstrated with millimetre-wave radar by using real aperture systems with narrow beams to increase antenna gain and received signal power. The All-weather Volcano Topography Imaging Sensor (AVTIS), which operates at 94 GHz \cite{Macfarlane2013}, was designed for this purpose and acquires a coarse resolution data cube by scanning across a scene of interest. Using a simple range to maximum received power algorithm, AVTIS has been successfully used to map volcanic lava dome growth since its first deployment in 2004 \cite{Wadge2005,Macfarlane2006,Macfarlane2013}. Results from field trials at the Soufri\`{e}re Hills volcano in Montserrat illustrated the potential of AVTIS for measuring lava dome topography at ranges of up to 3.8 km \cite{Wadge2005} and volcanic lava extrusion rates \cite{Wadge2008,Macfarlane2013}. Macfarlane et al. \cite{Macfarlane2006} further demonstrated the radar capabilities by mapping topographic changes at Arenal Volcano in Costa Rica alongside measurements of radar backscatter to examine an active lava flow. Despite these successful applications of AVTIS for volcano monitoring, the current surface elevation extraction methodology is simple and requires improvement in order to make full use of the information contained within the data cube.

% AIMS & PAPER SUMMARY
The primary aim of this study is to develop and demonstrate a new technique (the \textit{PCFilt-94} algorithm) to extract and filter 3D point clouds from coarse resolution $\left(R,\theta,\phi\right)$ radar data cubes. The accuracy of the new algorithm is evaluated by comparing 94 GHz radar 3D point clouds extracted using the \textit{PCFilt-94} algorithm with a validation point cloud acquired using TLS with centimetre accuracy. In Section \ref{Related_Work} we discuss previous studies investigating point cloud extraction and filtering methodologies. In Section \ref{Section_PCFilt-94}, we introduce the new \textit{PCFilt-94} algorithm, whilst in Section \ref{Section_Methods}, we describe field data collection and methods to quantify point cloud uncertainties. Section \ref{Section_Results} describes the performance and accuracy of the new methodology, whilst the results are discussed in Section \ref{Section_Discussion} before concluding in Section \ref{Section_Conclusion}. 

\section{Related Work} \label{Related_Work}
% RADAR DATA CUBES
The angular resolution of a radar is coarser than electro-optical instruments, hence their 3D mapping products are less accurate. Therefore, techniques to extract 3D surfaces from radar data cubes are generally more simplistic compared to methodologies developed for TLS systems. Macfarlane et al. \cite{Macfarlane2013} implemented a simple three-stage processing workflow to extract 3D point clouds from AVTIS data cubes (hereafter called the 'original algorithm'). The methodology used a two-way low pass filter to smooth range profiles of radar backscatter along each Line of Sight (LoS) (hereafter called a waveform), calculate the range to maximum radar received power along each waveform, and remove low power points by manually thresholding the resulting histogram of radar backscatter. The smoothing caused range migration and the terrain detection was sensitive to speckle which altered the apparent position of the strongest scattering target along LoS. Target detection based on Constant False Alarm Rate (CFAR) approaches have also been developed \cite{Foessel2001} but these methods are unsuitable for long-range mapping when terrain returns have a low Signal-to-Noise Ratio (SNR). Other methods have mapped 3D surfaces using statistically significant recurring terrain returns \cite{Weidinger2020} and simple first echo detection in pulsed radar systems \cite{Brooker2006}. The simplicity of the current set of terrain extraction techniques from radar data cubes does not account for the wide beams of radar systems and the impact of speckle, both of which may lead to erroneous returns within the terrain extraction methodology.

% LIDAR
The range to a target in TLS data cubes was traditionally measured as the first or last return along each waveform \cite{Baltsavias1999,Mallet2009} and so full-waveform Lidar/TLS data was developed to improve terrain extraction algorithms by digitising the full backscatter response of the target \cite{Mallet2009}. This has led to the development of more sophisticated algorithms to extract point clouds of surfaces. Decomposition of a waveform into its component Gaussian parts has been used to detect multiple terrain returns and isolate the existence of weak echoes \cite{Wagner2006}. Whilst this method assumes that backscattered echoes are normally distributed, they perform well in complex topographic environments. Alternatively, stacking of neighbouring backscatter waveforms that are scaled based on their position and relative intensities \cite{Teo2018} can be used to improve the waveform Signal-to-Noise Ratio (SNR). This methodology has been applied to TLS waveforms to detect weak echoes \cite{Yao2010}, enabling the extraction of denser 3D point clouds. Finally, deconvolution filters have been used to remove the beam pattern characteristics of the sensor from the backscattered signal in order to reduce noise and improve the positional accuracy of resultant point clouds \cite{Jutzi2006}. The more sophisticated TLS point cloud extraction methodologies have the potential to be transfered to radar data cubes for improved surface mapping in obscuring weather conditions.

% SPATIAL POINT CLOUD OUTLIER REMOVAL
3D point cloud filtering aims to remove non-terrain points and outliers from the point cloud. Simple methods based on thresholding the distribution of backscattered power values are typically used to remove the majority of non-terrain points from a point cloud \cite{Hofle2013,Macfarlane2013} and typically act as an initial processing step. Subsequent filtering steps aim to remove point cloud outliers that were not removed in the first stage and the detection of these points is usually based on the statistical, distance, and projection relationships between different points \cite{Han2017}. For example, Schall et al. \cite{Schall2005} identified point cloud clusters related to the presence of terrain using kernel density estimation, where outliers were identified as existing outside of terrain point clusters. Point cloud outlier detection using density estimators performs well on dense point clouds from TLS sensors but is less suitable for coarse resolution radar data cubes where the distance between points may be multiple metres, potentially leading to the false identification of outliers in a radar point cloud. Further, neighbourhood functions that compute the average distance of a point to it's neighbour have been used to identify point cloud clusters related to terrain \cite{Girardeau2005}, but this technique require manual intervention to determine a suitable threshold for a point to be considered an outlier.

\section{The PCFilt-94 Algorithm} \label{Section_PCFilt-94}
\begin{figure}[!t]
	\centering
	\includegraphics[width=\linewidth]{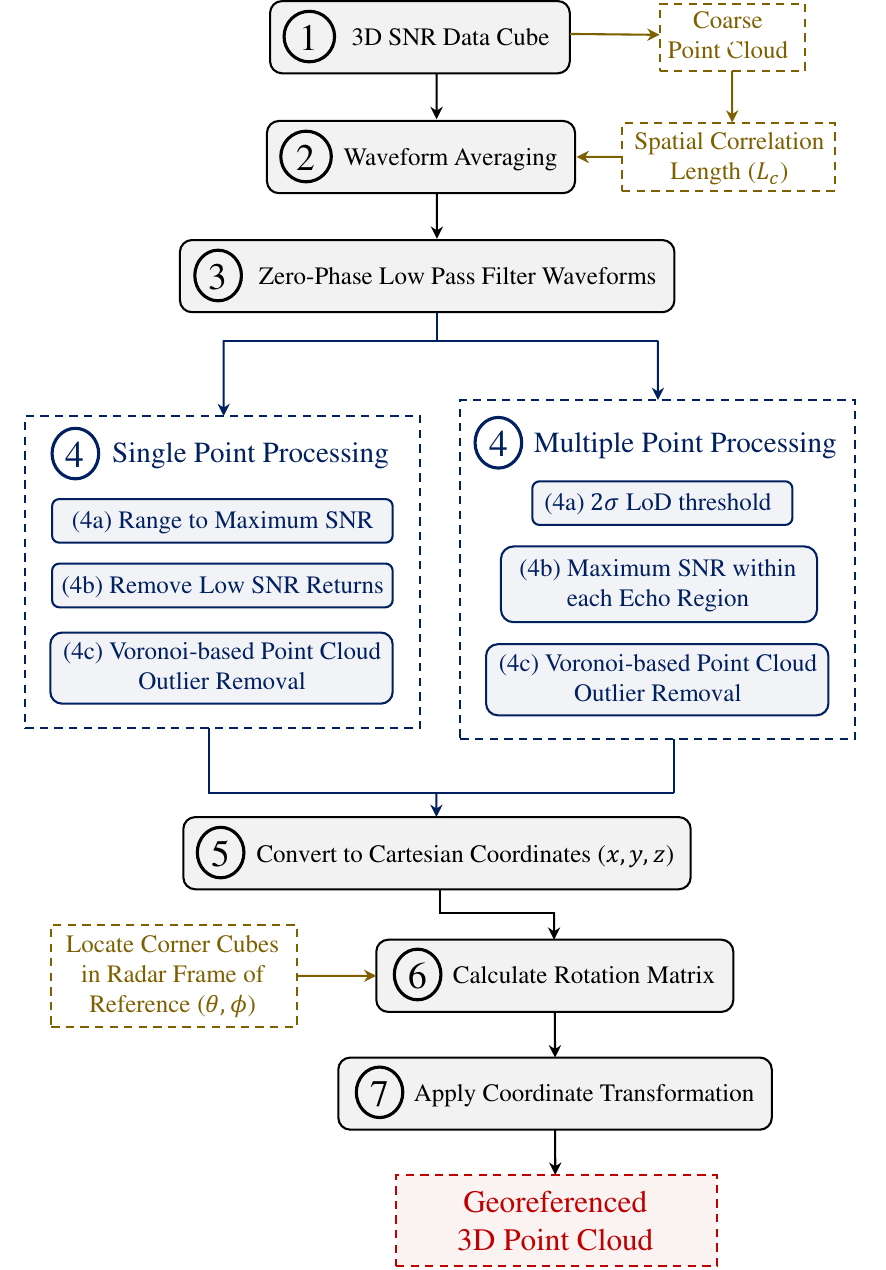}
	\caption{The \textit{PCFilt-94} processing chain used to derive 3D point clouds from radar data cubes. The box colours represent the following: grey = pre- and post-processing, blue = point cloud extraction and filtering, brown = mid-processing step.}	
	\label{Point_Cloud_Processing_Workflow}
\end{figure}

% INTRODUCTION TO CHAPTER
In this section, we introduce the \textit{PCFilt-94} algorithm which extracts and filters 3D point clouds from coarse data cubes acquired from 94 GHz radar systems. An overview of the new methodology is presented in Fig. \ref{Point_Cloud_Processing_Workflow}. Here, we provide the theoretical basis for the new methodology and use examples from data acquired using the 2\textsuperscript{nd} generation AVTIS (AVTIS2) radar to demonstrate its application.

\subsection{Theory}
\begin{figure}[!t]
	\centering
	\includegraphics[width=\linewidth]{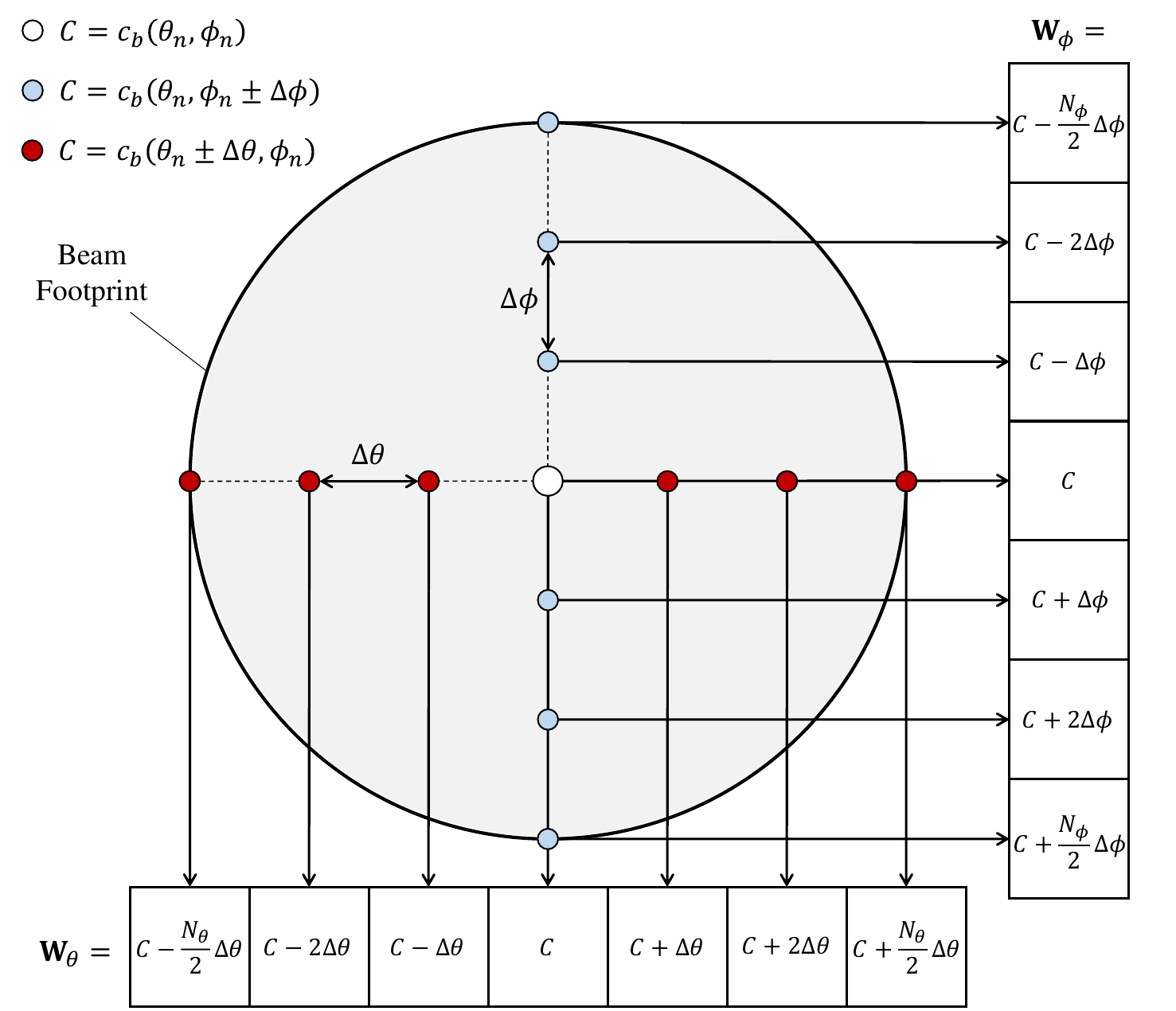}
	\caption{Schematic of a beam footprint area with a centre of $c_b{\left(\theta_n,\phi_n\right)}$ overlapping with footprints whose centre points are given by $c_b{\left(\theta_n \pm n_{\theta}\Delta\theta,\phi_n\right)}$ in azimuth and $c_b{\left(\theta_n,\phi_n \pm n_{\phi}\Delta\phi\right)}$ in elevation, where $\Delta\theta$ and $\Delta\phi$ are the angular intervals in azimuth and elevation, respectively. For simplicity, points beyond the azimuth and elevation axes are not shown. The vectors $\mathbf{W}_{\theta}$ and $\mathbf{W}_{\phi}$ are the set of members corresponding to those points overlapping with the footprint under interest. The substitution $C=c_b{\left(\theta_n,\phi_n\right)}$ is used due to space limitations.}	
	\label{Beam_Overlap}
\end{figure}

% BEAM OVERLAP
A real-aperture radar with a circular beam illuminates the terrain at sequential angular increments, where the centre of the footprint is given by $c_b{\left(\theta_n,\phi_n\right)}$ (Fig. \ref{Beam_Overlap}). Sequential beam footprints spatially overlap the scene and are thus at least partially correlated \cite{Ulaby2019}. The footprint centre at subsequent azimuthal and elevation steps $\left(c_b{\left(\theta_n,\phi_n\right)}\right)$ is then a function of the spatial sampling interval in azimuth $\left(\Delta\theta\right)$ and elevation $ \left(\Delta\phi\right)$:
\begin{align}
c_b{\left(\theta_n,\phi_n\right)} \approx c_b{\left(\theta_n \pm n_{\theta}\Delta\theta,\phi_n \pm n_{\phi}\Delta\phi \right)}
\end{align}
where $n_{\theta}$ and $n_{\phi}$ represent the number of subsequent angular increments in azimuth and elevation, respectively, within a radar scan and are both equal to 0 at the first radar angular position $\left(c_1\right)$. Two separate beam footprints along the azimuth and elevation axes, respectively, are considered to be overlapping when the distance between their respective beam centres is at least half the two-way radar beamwidth i.e. they are over-sampled. This is because Nyquist sampling (spatially fully sampled) requires points be acquired at every half two-way beamwidth. Therefore, the number of overlapping footprints across the full beam footprint (i.e. in the positive and negative azimuthal and elevation directions) can be calculated by:
\begin{align}
N_{\theta} = \dfrac{\theta_2}{\Delta{\theta}} ,\ N_{\phi} = \dfrac{\phi_2}{\Delta{\phi}} \quad \text{ where } N_{\theta},N_{\phi}=\mathbb{Z}
\end{align}
where $\theta_2$ and $\phi_2$ are the two-way radar beamwidths in azimuth and elevation, respectively. Using these results, the neighbouring waveforms that are contained within a beam footprint at position $c_b{\left(\theta_n,\phi_n\right)}$ along the azimuth $\left(\mathbf{W}_{\theta}\right)$ and elevation $\left(\mathbf{W}_{\phi}\right)$ axes can be formulated as follows (Fig. \ref{Beam_Overlap}):
\begin{multline}
\mathbf{W}_{\theta} = \biggl\{
c_b{\left(\theta_n-\dfrac{N_{\theta}}{2}\Delta\theta,\phi_n\right)}  , \dotsc , c_b{\left(\theta_n-n_{\theta}\Delta\theta,\phi_n\right)} , \\
c_b{\left(\theta_n,\phi_n\right)} , c_b{\left(\theta_n+n_{\theta}\Delta\theta,\phi_n\right)}  , \\
\dotsc , c_b{\left(\theta_n+\dfrac{N_{\theta}}{2}\Delta\theta,\phi_n\right)}  \biggr\}
\end{multline}
\begin{multline}
\mathbf{W}_{\phi} = \biggl\{ 
c_b{\left(\theta_n,\phi_n-\dfrac{N_{\phi}}{2}\Delta\phi\right)}  , \dotsc , c_b{\left(\theta_n,\phi_n-n_{\phi}\Delta\phi\right)} , \\
c_b{\left(\theta_n,\phi_n\right)} , c_b{\left(\theta_n,\phi_n+n_{\phi}\Delta\phi\right)} , \\
\dotsc , c_b{\left(\theta_n,\phi_n+\dfrac{N_{\phi}}{2}\Delta\phi\right)} \biggr\}
\end{multline}
Each member of the $\mathbf{W}_{\theta}$ and $\mathbf{W}_{\phi}$ sets are defined by the centre of neighbouring footprints, hence the minimum overlap of the waveform centred on $ c_b{\left(\theta_n\pm\dfrac{N_{\theta}}{2}\Delta\theta , \phi_n\pm\dfrac{N_{\theta}}{2}\Delta\phi\right)} $ is half its beam area and therefore satisfies the Nyquist criterion for using neighbouring samples to reconstruct the signal at $c_b{\left(\theta_n,\phi_n\right)}$. For a radar transmitting a circular beam pattern, the angular coordinates of waveforms that exist beyond the azimuth $\left(\theta\right)$ and elevation $\left(\phi\right)$ axes are considered to be contained within the beam footprint at $c_b{\left(\theta_n,\phi_n\right)}$ $\left(\mathbf{w}_{\theta\phi} \right)$ under the following condition:
\begin{multline}
\left\{ \mathbf{w}_{\theta\phi} | d\left[ c_b{\left(\theta_n,\phi_n\right)} , c_b{\left(\theta_n \pm n_{\theta}\Delta\theta,\phi_n \pm n_{\phi}\Delta\phi \right)} \right] \right. \\
\left. < \dfrac{\beta{\left(\theta_2,\phi_2\right)}}{2} \right\}
\end{multline}
That is to say, the centre of a beam footprint given by $c_b{\left(\theta_n \pm n_{\theta}\Delta\theta,\phi_n \pm n_{\phi}\Delta\phi \right)}$ overlaps with $c_b{\left(\theta_n,\phi_n\right)}$ if the angular distance $\left(d\right)$ between the two is smaller than half the two-way radar beamwidth $\left(\beta{\left(\theta_2,\phi_2\right)}\right)$. Bringing all of the above together, the complete set of waveforms overlapping with any footprint $\left(c_b{\left(\theta_n,\phi_n\right)}\right)$ is given by:
\begin{align}
\mathbf{W}_{\theta\phi} = \left\{ \mathbf{W}_{\theta} \mathbf{W}_{\phi} \mathbf{w}_{\theta\phi} \right\}
\end{align}
These equations demonstrate that the characteristics of a signal at any particular angular position $c_b{\left(\theta_b,\phi_n\right)}$ estimated from a single waveform can be reconstructed by integrating those contained within the beam overlap area defined by the set $\mathbf{W}_{\theta\phi}$.

% BEAM OVERLAP Fig.

\subsection{Waveform Averaging}
\begin{figure}[!t]
	\centering
	\includegraphics[width=\linewidth]{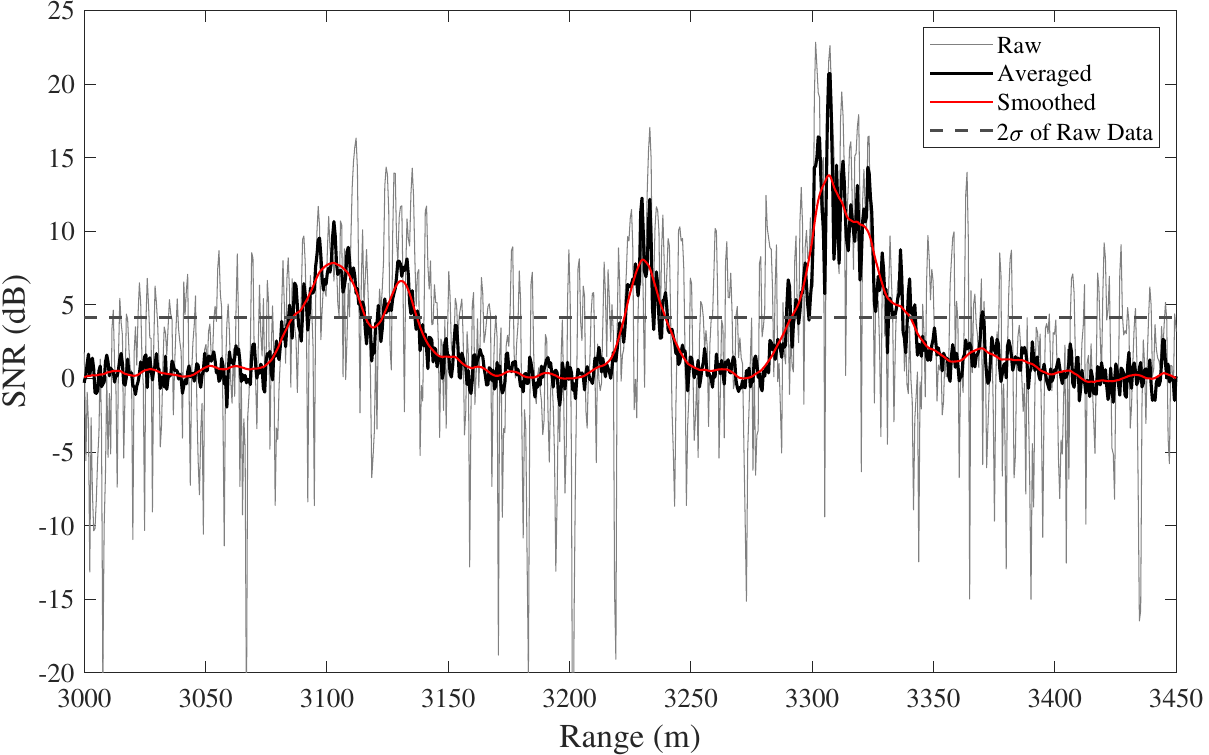}
	\caption{Subset of an AVTIS2 waveform showing the improvement in SNR when neighbouring waveforms are averaged.}	
	\label{Waveform_Stacking}
\end{figure}

% SPATIAL SAMPLING
We develop a method based on averaging neighbouring waveforms from locally surrounding lines of sight to suppress noise along a raw Signal-to-Noise Ratio (SNR) waveform (Fig. \ref{Waveform_Stacking}). This spatial averaging smooths the impact of rapid coherent fluctuations due to speckle and reveals a smoother underlying pattern. In theory, the terrain signal SNR can be improved by non-coherently averaging all waveforms (i.e. frequency spectra) from $\mathbf{W}_{\theta\phi}$ to generate an improved estimate of the range to footprint centre $\left(C_b\right)$:
\begin{align}
C_b{\left(\theta_n,\phi_n\right)} = \dfrac{1}{N} \sum_{N=1}^{N} \mathbf{W}_{\theta\phi}
\end{align}
where $N$ is the number of members contained in $\mathbf{W}_{\theta\phi}$. However, the path length between the radar and a target at the centre of the beam footprint $\left(c_b{\left(\theta_n,\phi_n\right)}\right)$ changes as the radar scans across the terrain to a different angular position, resulting in increasing signal decorrelation as the radar moves further away from the target scattering centre. This fading decorrelation distance $\left(L_d\right)$ is proportional to the size of the target, which for distributed, beam-filling terrain can be approximated by the diameter spot size size in azimuth $\left(\theta\right)$ and elevation $\left(\phi\right)$ \cite{Richards2010}:
\begin{align}
L_d{\left(\theta\right)} = \dfrac{\lambda}{2R\tan{\theta_2/2}} , \quad L_d{\left(\phi\right)} = \dfrac{\lambda}{2R\tan{\phi_2/2}}
\end{align}
where $\theta_2$ and $\phi_2$ are the two-way beamwidths in azimuth and elevation, respectively, and $\lambda$ is the wavelength. Conversely, the degree of spatial correlation for randomly rough surfaces can be calculated using the correlation length $\left(L_c\right)$ with a Gaussian correlation function \cite{Rees2005,Ulaby2014}:
\begin{align}
\label{Equation_Correlation_Length}
L_c = \sqrt{2} \dfrac{\sigma_h}{\sigma_m}
\end{align}
where $\sigma_h$ is the root mean square (rms) height deviation and $\sigma_m$ is the rms slope. It follows that any member of the set $\mathbf{W}_{\theta\phi}$ is statistically independent from $c_b{\left(\theta_n,\phi_n\right)}$ when $L_c<L_d$. On this basis, $C_b{\left(\theta_n,\phi_n\right)}$ is estimated under the following conditions:
\begin{align}
\label{Waveform_Averaging_Equation}
C_b{\left(\theta_n,\phi_n\right)} =
\begin{cases}
\dfrac{1}{N} \sum\limits_{N=1}^{N}{\mathbf{W}_{\theta\phi}} ,& \text{if } L_c \geq L_d \\
c_b{\left(\theta_n,\phi_n\right)} ,& \text{if } L_c < L_d
\end{cases}
\end{align}
The waveform averaging method is therefore simply the application of Equation \ref{Waveform_Averaging_Equation} to a 3D $\left(R,\theta,\phi\right)$ data cube. It is worth noting that typical values for $L_c$ over terrain are $>$0.4 m and over sky are $<$0.2 m, hence the condition $L_c<L_d$ is almost never met over terrain but almost always met when scanning the sky.

% WAVEFORM STACKING EXAMPLE

% GEOSPATIAL WAVEFORM STACKING: PRACTICE
The value of $L_c$ at each angular position $\left(c_b{\left(\theta_n,\phi_n\right)}\right)$ is estimated from a coarse point cloud extracted from the original $\left(R,\theta,\phi\right)$ data cube by applying a zero-phase low pass filter to each waveform \cite{Gustafsson1996} and applying the range to maximum SNR algorithm \cite{Macfarlane2013}. The coarse $\left(R,\theta,\phi\right)$ point cloud in spherical coordinates $\left(R_n,\theta_n,\phi_n\right)$ is then converted to a Cartesian frame of reference with the radar at the origin $\left(O\right)$ (see section \ref{Georeferencing}). The rms height deviation $\left(\sigma_h\right)$ and rms slope $\left(\sigma_m\right)$ at each angular position in a radar scan is calculated by extracting all surrounding points lying within a diameter of the two-way radar beamwidth $\left(c_{\text{nn}}\right)$. For these points, $\sigma_h$ of $c_{\text{nn}}$ is calculated by \cite{Rees2005,Richards2010,Ulaby2014}:
\begin{align}
\sigma_{h} = \sqrt{ \dfrac{1}{N} \sum_{N=1}^{N} \left[ c_{\text{nn}}\left(z\right) - {\left\langle c_{\text{nn}}\left(z\right) \right\rangle} \right]^2 }
\end{align}
where $c_{\text{nn}}\left(z\right)$ represents each sampled height from the extracted points, ${\langle c_{\text{nn}}\left(z\right) \rangle}$ represents the mean surface height, and $N$ the number of points contained in $c_{\text{nn}}$. Similarly, the rms surface slope $\left(\sigma_m\right)$ is also calculated for $c_{\text{nn}}$ using:
\begin{align}
\sigma_{m} = \sqrt{ \dfrac{1}{N} \sum_{N=1}^{N} \left[ \dfrac{{\partial}^2}{\partial{x}\partial{y}} c_{\text{nn}} - \left\langle \dfrac{{\partial}^2}{\partial{x}\partial{y}} c_{\text{nn}} \right\rangle \right]^2 }
\end{align}
where ${\partial}^2{c_{\text{nn}}}/\partial{x}\partial{y}$ represents the local surface slopes in the $x$ and $y$ directions, and $\langle {\partial}^2{c_{\text{nn}}}/\partial{x}\partial{y} \rangle$ represents the mean slopes in $x$ and $y$. The values of $\sigma_h$ and $\sigma_m$ are then used to calculate the correlation length from Equation \ref{Equation_Correlation_Length} at each angular position in the radar data cube. Under the condition $L_c \geq L_d$, the waveform averaging technique suppresses noise and increases SNR as shown in Fig. \ref{Waveform_Stacking} (black line). The zero-phase low pass filter \cite{Gustafsson1996} is now applied to the averaged waveform (black line in Fig. \ref{Waveform_Stacking}) using a sliding window equal to the number of waveforms averaged at each angular position.

\subsection{Multiple Target Detection}
% ZERO-PHASE FILTERING, LoD, AND TARGET DETECTION.
In the example shown in Fig. \ref{Waveform_Stacking}, the improved SNR reveals the existence of three bulk targets and applying the range to maximum SNR algorithm (single-point processing) extracts the strongest returned signal at $\sim$3,300 m. In the new algorithm, we extract additional returns (e.g. at $\sim$3,100 m and $\sim$3,225 m) along the averaged, smoothed waveform (red line in Fig. \ref{Waveform_Stacking}) where they exceed two standard deviations $\left(2\times\sigma\right)$ of the full SNR averaged waveform (black line in Fig. \ref{Waveform_Stacking}). This threshold was chosen because terrain returns may represent anywhere between 50 to 500 range bins (0.61-6\%) across a waveform, hence there is a high probability ($>$95\%) that terrain returns will exist if they exceed the $2\sigma$ threshold. In the example shown in Fig. \ref{Waveform_Stacking}, four individual terrain echoes are isolated along the waveform and the range to each surface is estimated based on the range to maximum SNR within each local region. This process will extract noise when no terrain returns are present along a waveform and these points are removed during point cloud filtering (see next section). This new point cloud extraction methodology (multiple-point processing) increases the resultant point cloud density and also enables multiple surfaces along a Line of Sight (LoS) to be extracted, which was not possible using the range to maximum SNR approach. 

\subsection{SNR Point Cloud Filtering}
\begin{figure}[!t]
	\centering
	\includegraphics[width=\linewidth]{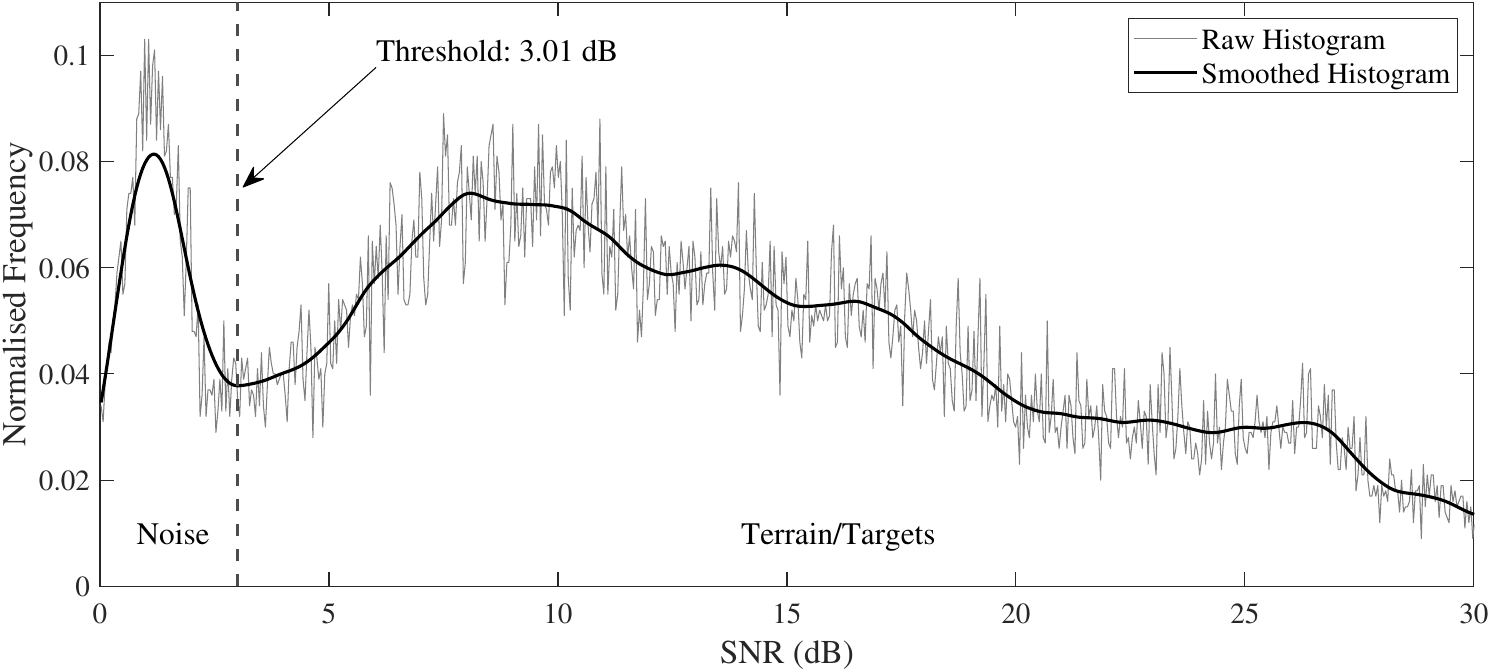}
	\caption{Example of the low SNR point removal technique. The histogram contains 1,000 bins and is typical for AVTIS2 point clouds scanning terrain in an upwards direction with sky returns at high elevation angles.}	
	\label{Low_SNR_Removal}
\end{figure}

\begin{figure}[!t]
	\centering
	\includegraphics[width=\linewidth]{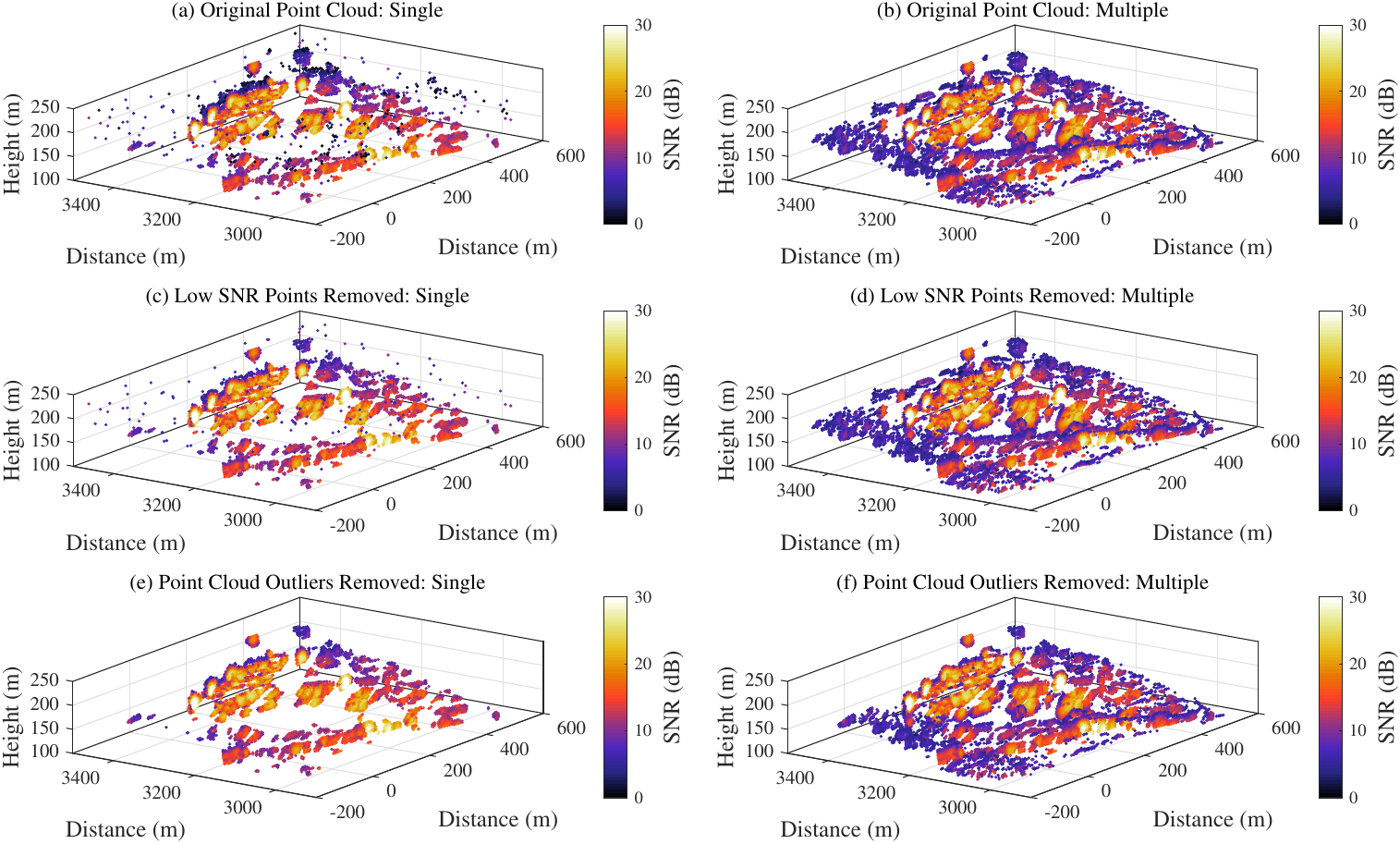}
\caption{Effect of each filtering step on the removal of points in an AVTIS2 3D point cloud. (a) Original point cloud (Single), (b) Original point cloud (Multiple), (c) low SNR returns removed (Single), (d) low SNR returns removed (Multiple), (d) spatial outliers removed based on Voronoi diagrams (Single), and (e) spatial outliers removed based on Voronoi diagrams (Multiple). Data is taken from AVTIS2 measurements of Balmullo Quarry at Site 1 on 6 February 2014.}	
	\label{PointCloud_Filtering_Example}
\end{figure}

% SNR Filtering
For single-point processing, the first stage of filtering involves removing low SNR points (Fig. \ref{Point_Cloud_Processing_Workflow}), which result from the radar scanning across a low scattering surface/medium such as smooth water or the sky. The reduced backscatter in these locations results in the range to maximum SNR algorithm measuring random fluctuations in the noise floor rather than terrain, yielding the isolated outlier points shown in the original point cloud in Fig. \ref{PointCloud_Filtering_Example}a and Fig. \ref{PointCloud_Filtering_Example}b. A typical histogram of waveform-averaged SNR values with 1,000 bins is plotted in Fig. \ref{Low_SNR_Removal}, illustrating a broad range of terrain returns above ${\sim}5$ dB and a well-defined distribution of low SNR values below ${\sim}3$ dB representing Rayleigh distributed noise \cite{Ulaby2014,Richards2010}. Low SNR points are removed based on this histogram by first applying a smoothing function using a weighted regression sliding window (i.e. the locally weighted scatterplot smoothing, or lowess, method; Fig. \ref{Low_SNR_Removal}). The sliding window nominally has a length of 5\% of the histogram size, which represents 50 bins for a 1,000 bin histogram. The lowess curve fitting routine is effective at removing high frequency signal variability \cite{Cleveland1979} and is thus suitable for the purposes of smoothing the SNR histogram containing signals from radar noise and terrain. The first trough in the smoothed histogram is used as a threshold to remove low SNR returns (Fig. \ref{Low_SNR_Removal}) and the impact on the point cloud is shown in Fig. \ref{PointCloud_Filtering_Example}c and Fig. \ref{PointCloud_Filtering_Example}d. This trough is detected at the point in the positive infinity direction when the curve gradient switches from negative to positive. This step is not applied to the multiple-point processing methodology (Fig. \ref{Point_Cloud_Processing_Workflow}) as the $2\sigma$ threshold avoids low noise power returns; point cloud outliers are instead removed in the spatial filtering step discussed next.

% SNR FILTERING Fig.

% POINT CLOUD FILTERING EXAMPLE

\subsection{Spatial Point Cloud Filtering}
\begin{figure}[!t]
	\centering
	\includegraphics[width=\linewidth]{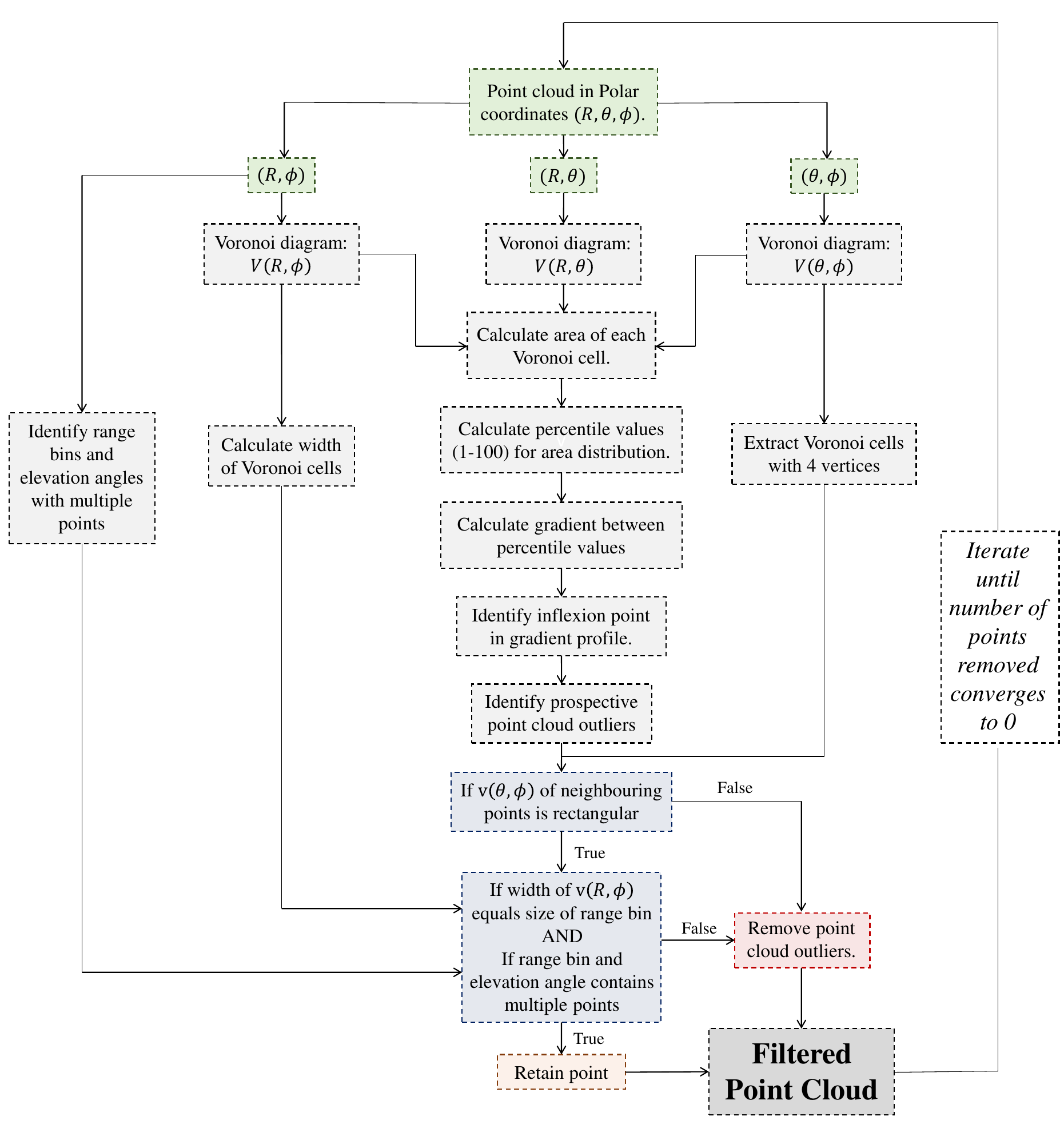}
	\caption{Flow chart of the Voronoi point cloud outlier detection algorithm.}	
	\label{Voronoi_Workflow}
\end{figure}

\begin{figure}[!t]
	\centering
	\includegraphics[width=\linewidth]{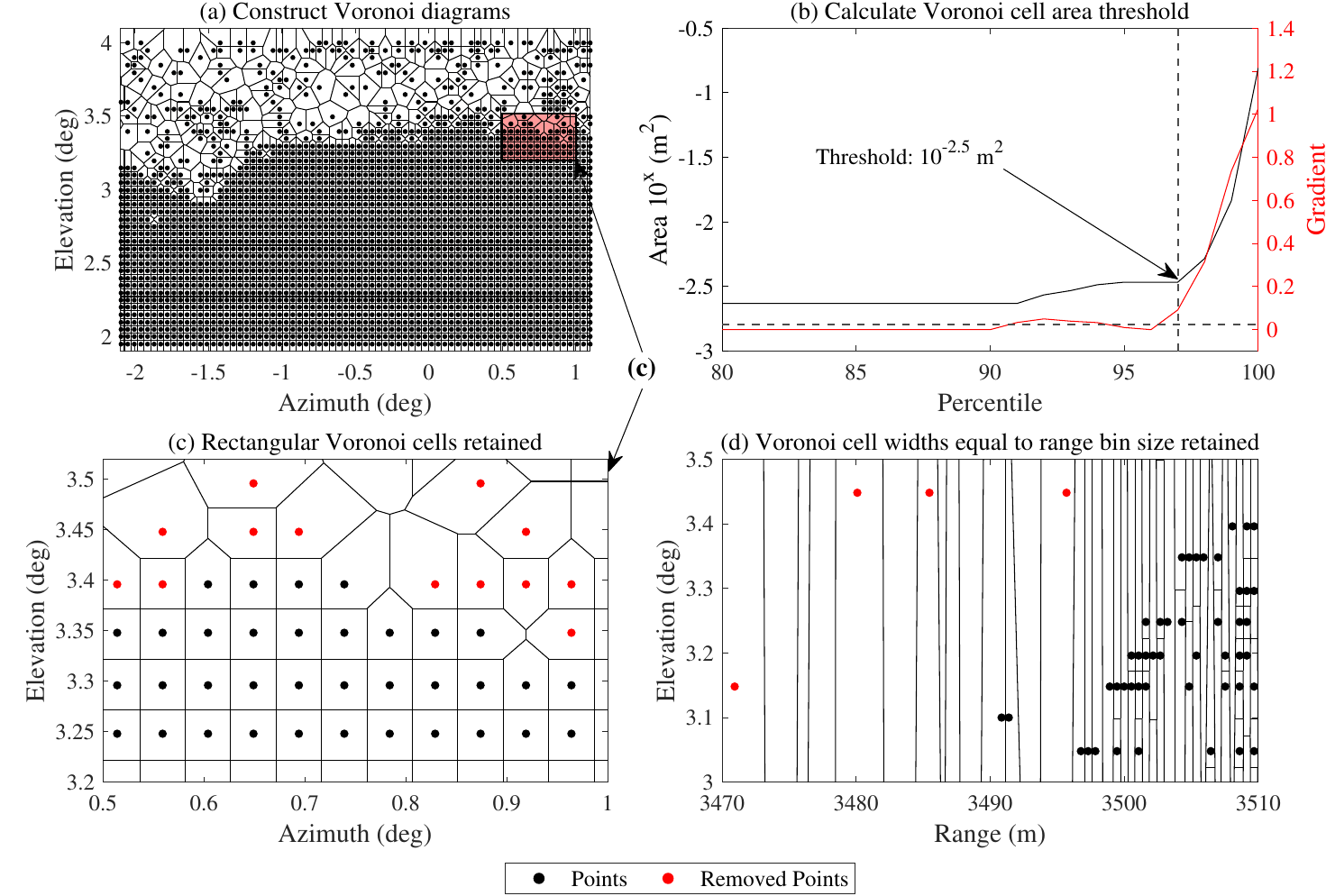}
	\caption{Example of Voronoi filtering on a point cloud after removal of low SNR returns. (a) Original point cloud (black points) overlaid with a Voronoi diagram (black lines). Large Voronoi cells bound the locations of point cloud outliers. (b) Percentile values for the Voronoi area distribution between 1 and 100 (black), plotted alongside the gradient along the line (red). Horizontal dashed line is the gradient mean, whilst the vertical dashed line represents the percentile at which the gradient line exceeds this value and continues to increase. This is used to calculated the threshold on the area percentile distribution. (c) Example of a section of the point cloud (red box in panel (a)) where terrain points are clustered and bounded by rectangular Voronoi cells. (d) Removal of points bounded by Voronoi cells that are larger than the size of the range bin size.}	
\label{Voronoi_Example}
\end{figure}

% VORONOI BASED OUTLIER REMOVAL
Spatial outliers that persist after the SNR filtering stage are randomly distributed in space, hence a method to identify and remove these `non-terrain' points based on Voronoi diagrams was developed, see Fig. \ref{Voronoi_Workflow} for a summary. Voronoi diagrams are created by computing the Euclidean distance of each point on a plane $\left(p\right)$ to a point in the point cloud $\left(p_i\right)$ and a line drawn at locations where the distance between two or more points is identical \cite{Okabe2017}. Joining multiple lines together creates a Voronoi cell $\left(\text{v}{\left(p_i\right)}\right)$ for a specific point in a point cloud $\left(p_i\right)$ and can be defined as:
\begin{align}
\text{v}{\left(p_i\right)} = \left\{ p | d{\left(p,p_i\right)} \leq d{(p,p_j)}, j \neq i, j=\left[1,\ldots,n_j\right] \right\}
\end{align}
where $d{\left(p,p_i\right)}$ represents the Euclidean distance between $p$ and $p_i$, $d{\left(p,p_j\right)}$ represents the Euclidean distance between $p$ and $p_j$, and $n_j$ represents the number of points in the point cloud. $p_j$ is another arbitrary point in the point cloud. Lines are mapped along the coordinates of the plane where $d\left(p,p_i\right) = d\left(p,p_j\right)$ which bound the Voronoi cell. Grouping these Voronoi cells together creates a Voronoi diagram derived from the point cloud $\left(V{\left(P\right)}\right)$:
\begin{align}
V(P) = \left\{ \text{v}{\left(p_i\right)}, \ldots, \text{v}{\left(p_n\right)} \right\}
\end{align}
where $n$ is the number of cells in the Voronoi diagram and is equal to the number of points in the point cloud. Here, the Voronoi diagram is computed in 2D for the three point cloud geometries: $(R,\theta)$, $(R,\phi)$, and $(\theta,\phi)$ (Fig. \ref{Voronoi_Example}a). The area of each Voronoi cell within each geometry is then calculated after which the percentile value of the area distribution between 1 and 100 is computed. Voronoi cell areas surrounding point cloud outliers are expected to be clustered towards the upper percentile region because: 1) they are assumed to be much larger compared to those surrounding terrain points, and 2) represent a small contribution to the total number of points in a point cloud. This is represented by an inflexion point along the 1D percentile distribution (Fig. \ref{Voronoi_Example}b) and is detected where the gradient of the 1D percentile line exceeds the mean of the gradient profile and continues to increase (Fig. \ref{Voronoi_Example}b). Points bounded by a Voronoi cell whose area is above this threshold are considered outliers.

% VORONOI FILTERING WORKFLOW

% VORONOI FILTERING EXAMPLE

% IDENTIFYING AND REMOVING OUTLIERS
Points identified as outliers are removed when the following condition is met:
\begin{enumerate}
\item{\textit{The Voronoi cells of one or fewer neighbouring points in the azimuth and elevation dimensions {\normalfont$\left(\text{v}{\left(\theta,\phi\right)}\right)$} is rectangular}: Because a point cloud in spherical coordinates is sampled at regular increments in azimuth $\left(\theta\right)$ and elevation $\left(\phi\right)$, the distance between terrain points is equal and the resulting Voronoi diagrams bounding terrain points are rectangular (Fig. \ref{Voronoi_Example}c). The clustered set of points representing terrain then becomes a collection of rectangular Voronoi cells whilst outliers are represented by irregular Voronoi cells. Therefore, point cloud outliers are identified when a maximum of one of its neighbours is bounded by a rectangular Voronoi cell.}
\end{enumerate}
If this condition is not met, the two following additional constraints must be met for the point to be considered an outlier:
\begin{enumerate}
\item[2a)]{\textit{The Voronoi cells of one or fewer neighbouring points is rectangular but the width of the Voronoi cell bounding the point in range and elevation {\normalfont$\left(\text{v}{\left(R,\phi\right)}\right)$} exceeds the size of the range bin}: Terrain points with coordinates $\left(R,\phi\right)$ can extend over multiple elevation angles but are restricted in the range direction by the size of the AVTIS2 range bin (Fig. \ref{Voronoi_Example}d). Therefore, $\left(\theta,\phi\right)$ points bounded by Voronoi cells whose width is larger than the size of the AVTIS2 range bin are classified as outliers.}
\item[2b)]{\textit{The Voronoi cells of one or fewer neighbouring points is rectangular but the point does not overlap with other points at the same elevation angle and range bin}: A point that may be considered terrain based on its rectangular Voronoi cell $\left(\text{v}{\left(\theta,\phi\right)}\right)$ will be classified as an outlier if it is the only point in the point cloud representing its elevation angle and range bin.}
\end{enumerate}
A point cloud outlier can therefore be identified where it meets condition (1) or where it meets both conditions (2a) and (2b). These outliers are subsequently removed from the point cloud.

\subsection{Georeferencing} \label{Georeferencing}
% GEOREFERENCING OVERVIEW
The filtered 3D point cloud in spherical coordinates $\left(\mathbf{C}_{\text{p}}\right)$ is next converted to a local cartesian frame of reference $\left(\mathbf{C}_{\text{Loc}}\right)$:
\begin{equation}
\mathbf{C}_{\text{Loc}} = \mathbf{R}_{\text{pc}} \mathbf{C}_{\text{p}}
\end{equation}
\begin{equation}
\label{SphericaltoCartesian_Matrix}
\begin{bmatrix}
 x \\
 y \\
 z 
\end{bmatrix}_{\text{Loc}}
=
\begin{bmatrix}
 \sin{\theta}\cos{\phi} & \cos{\theta}\cos{\theta} & -\sin{\phi} \\
 \sin{\theta}\sin{\phi} & \cos{\theta}\sin{\phi} & \cos{\phi} \\
 \cos{\theta} & -\sin{\theta} & 0 
\end{bmatrix}_{\text{pc}}
\begin{bmatrix}
R \\
\theta \\
\phi
\end{bmatrix}_{\text{p}}
\end{equation}
where $\mathbf{R}_{\text{pc}}$ is the spherical $\left(R,\theta,\phi\right)$ to cartesian $\left(x,y,z\right)$ rotation matrix. To obtain the point cloud coordinates in geographic space $\left(\mathbf{C_{\text{Geo}}}\right)$, a 3D coordinate transformation $\left(\mathbf{R}_{\text{lg}}\right)$ is applied to $\mathbf{C}_{\text{Loc}}$:
\begin{equation}
\label{Georeferencing_Equation}
\mathbf{C_{\text{Geo}}} = \mathbf{A} + \mathbf{R}_{\text{lg}} \mathbf{C}_{\text{Loc}}
\end{equation}
\begin{equation}
\begin{bmatrix}
 x \\
 y \\
 z 
\end{bmatrix}_{\text{Geo}}
= 
\begin{bmatrix}
 x_{\text{A2}} \\
 y_{\text{A2}} \\
 z_{\text{A2}} 
\end{bmatrix}
+
\mathbf{R}_{\text{lg}}
\begin{bmatrix}
 x \\
 y \\
 z 
\end{bmatrix}_{\text{Loc}}
\end{equation}
where $\mathbf{A}$ represents the radar position in geographic space. In this way, the coordinate transformation $\left(\mathbf{R}_{\text{lg}}\right)$ rotates the 3D point cloud around the radar position and translates it into a new geographic coordinate system using the geolocated coordinates of the radar $\left(\mathbf{A}\right)$. Here, the cartesian Earth-Centered, Earth-Fixed (ECEF) coordinate system is used which has an origin at the centre of the Earth. The coordinate transform  $\left(\mathbf{R}_\text{lg}\right)$ is computed by comparing the orientation of Ground Control Points (GCPs) relative to the radar in both local and ECEF coordinates. The cartesian coordinates of the GCP in local and ECEF coordinates (measured from dGPS) are then normalised by the slant range distance between the radar and the GCP. 

% ROTATION OF DATA
To obtain the 3 $\times$ 3 coordinate transformation (rotation) matrix $\left(\mathbf{R}_{\text{lg}}\right)$, Equation \ref{Georeferencing_Equation} can be expressed in terms of GCP coordinates centred on the radar as the origin:
\begin{equation}
\label{Alignment_Equation}
\mathbf{G}_{\text{rGeo}} = \mathbf{R} \mathbf{G}_{\text{Loc}}
\end{equation}
\begin{equation}
\begin{bmatrix}
 x \\
 y \\
 z 
\end{bmatrix}_{\text{rGeo}}
= \mathbf{R}
\begin{bmatrix}
 x \\
 y \\
 z 
\end{bmatrix}_{\text{Loc}}
\end{equation}
where $\mathbf{G}_{\text{rGeo}}$ represents the GCP orientation relative to the radar in ECEF (i.e. with no translation applied by the radar ECEF position $\left(\mathbf{A}\right)$) and $\mathbf{G}_\text{Loc}$ are the GCP locations in local cartesian coordinates. Multiplying both sides of Equation \ref{Alignment_Equation} by $\mathbf{G}^\mathbf{T}_{\text{rGeo}}$ rotates and stretches the resulting matrix:
\begin{equation}
\mathbf{R} \mathbf{G}_{\text{Loc}} \mathbf{G}^\mathbf{T}_{\text{rGeo}} = \mathbf{U}  \mathbf{\Sigma}
\end{equation}
Rearranging gives:
\begin{equation}
\label{SVD}
\mathbf{G}_{\text{Loc}} \mathbf{G}^\mathbf{T}_{\text{rGeo}} = \mathbf{U}  \mathbf{\Sigma} \mathbf{R}^\mathbf{T}
\end{equation}
Equation \ref{SVD} demonstrates that the matrix $\mathbf{G}_{\text{Loc}} \mathbf{G}^\mathbf{T}_{\text{rGeo}}$ can be decomposed into two rotation matrices $\mathbf{U}$ and $\mathbf{R}^\mathbf{T}$, and a stretching or shrinking factor represented by $\mathbf{\Sigma}$. This is known as Singular Value Decomposition (SVD) and can be computed in most software packages e.g. Matlab. The left hand side of Equation \ref{SVD} represents the covariance matrix between $\mathbf{G}_{\text{Loc}}$ and $\mathbf{G}_{\text{rGeo}}$ and accumulates the combined rotation across the three axis in $x$, $y$, and $z$. Given that both $\mathbf{G}_{\text{Loc}}$ and $\mathbf{G}_{\text{rGeo}}$ are unit vectors, $\mathbf{\Sigma}$ can be ignored. Therefore Equation \ref{SVD} can be rewritten as:
\begin{align}
\mathbf{G}_{\text{rGeo}} = {\left( \mathbf{U} \mathbf{R}^\mathbf{T} \right)}^\mathbf{T} \mathbf{G}_{\text{Loc}} 
\end{align}
and therefore:
\begin{align}
\mathbf{R}_{\text{lg}} = {\left( \mathbf{U} \mathbf{R}^\mathbf{T} \right)}^\mathbf{T}
\end{align}
Overall, the coordinate transformation $\mathbf{R}_{\text{lg}}$ is obtained from the SVD of the covariance matrix between the GCPs in local $\left(\mathbf{G}_{\text{Loc}}\right)$ and ECEF $\left(\mathbf{G}_{\text{rGeo}}\right)$ coordinates. $\mathbf{R}_{\text{lg}}$ can now been used to georeference radar point clouds into ECEF coordinates using Equation \ref{Georeferencing_Equation}. Increasing the number of GCPs used to estimate the covariance matrix will increase the accuracy of $\mathbf{R}_{\text{lg}}$ up to a threshold number of GCPs, beyond which the accuracy will converge with no significant impact on georeferencing performance. The impact of using different combinations of GCPs is analysed further in section \ref{4_Georeferencing_Performance}. 

\section{Methods} \label{Section_Methods}
\subsection{Study Site and Instruments} \label{Section_StudySite_Instruments}
\begin{figure}[!t]
	\centering
	\includegraphics[width=\linewidth]{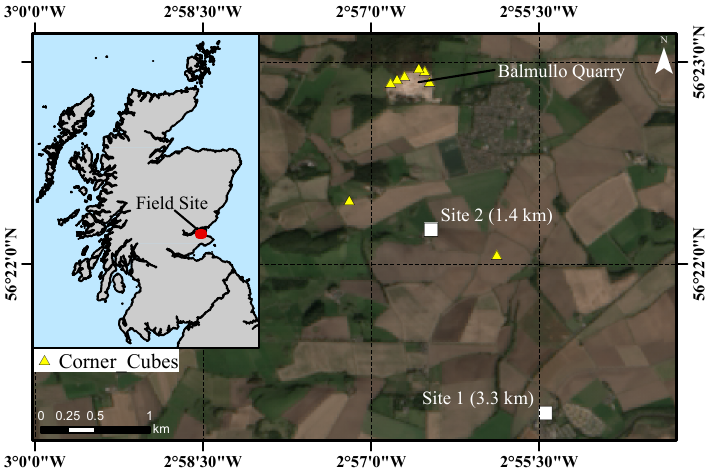}
	\caption{Map showing the study sites, Balmullo Quarry and the locations of the GCPs.}	
	\label{Balmullo_Quarry_Map}
\end{figure}

% BALMULLO QUARRY FIELD SITE
The topography of Balmullo Quarry in Scotland (Fig. \ref{Balmullo_Quarry_Map}) was surveyed on 6 and 7 February 2014 using the AVTIS2 94 GHz radar and a short-range TLS (Table \ref{Balmullo_Instrument_Summary}). 
Balmullo Quarry is approximately 300 m $\times$ 450 m in size and has been excavated out from the side of a small hill. Its surface consists of bare rock and rubble, whilst isolated patches of vegetation can be found around its margin. The lack of vegetation across the quarry face makes it an ideal location to compare point clouds between millimetre-wave radar and TLS as we can assume each sensor will be mapping the same surface horizon i.e. there will be no signal penetration. However, the topography across the quarry is complex, with steep rock faces interspersed with horizontal vehicle access benches that lead to sharp discontinuities in the 3D shape of the quarry.

% SHORT-RANGE TLS AND AVTIS2
\begin{table}[!t]
\caption{Specifications of the two sensors used in this study: AVTIS2 and the Leica ScanStation C10 short-range TLS.}
\label{Balmullo_Instrument_Summary}
\centering
\begin{tabular}{|c|c|c|c|c|}
\hline
\textbf{Parameter} & \textbf{\shortstack{Millimetre-wave\\Radar}} & \textbf{\shortstack{Short-range\\TLS}} \\
\hline
System & AVTIS2 & Leica ScanStation C10 \\
\hline
Measurement Type & FMCW & Pulsed \\
\hline
Range Resolution & \shortstack{Site 1: 0.54 m \\ Site 2: 0.5 m} & 0.07 m \\
\hline
Centre Frequency & 94 GHz & 564 THz \\
\hline
Wavelength & 3.19 mm & 532 nm \\
\hline
Two-way Beamwidth & \shortstack{Azimuth $\left(\theta\right)$: = 0.33$\degree$ \\ Elevation $\left(\phi\right)$: 0.35$\degree$} & 0.0041$\degree$ \\
\hline
Max Range & \shortstack{Site 1: 4,408 m \\ Site 2: 4,096 m} & 300 m \\
\hline
\end{tabular}
\end{table}

% BALMULLO QUARRY ANNOTATED DIAGRAM

\begin{figure}[!t]
	\centering
	\includegraphics[width=\linewidth]{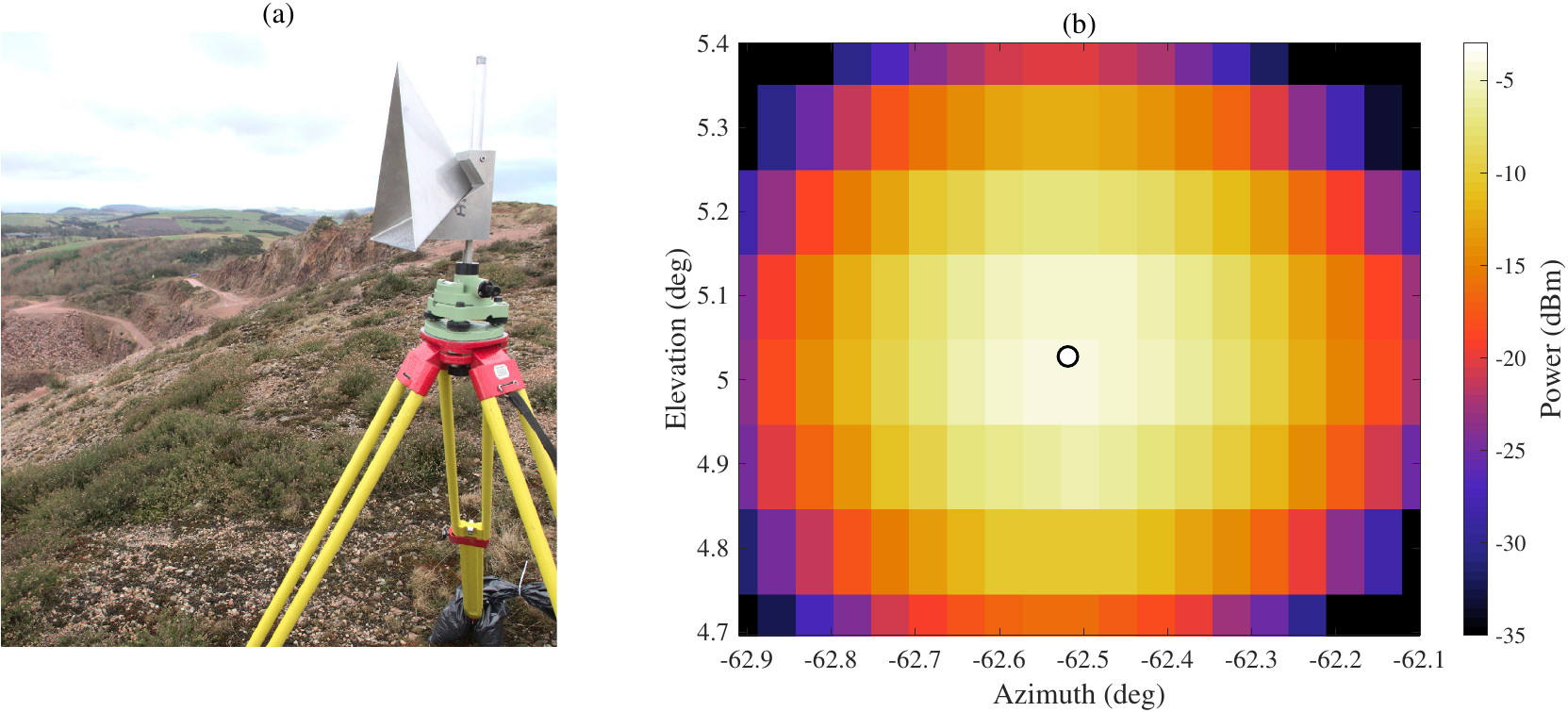}
	\caption{(a) Trihedral corner reflector used as a Ground Control Point (GCP) and (b) an example AVTIS$2$ scan of a trihedral corner reflector. The white circle indicates the centre of the GCP.}	
	\label{CC_Georeferencing}
\end{figure}

% FIELD SET-UP
AVTIS2 was deployed at two locations $\sim$3.3 km (Site 1) and $\sim$1.4 km (Site 2) from the quarry to assess its mapping capabilities at different ranges. The short-range TLS scanned the same region from one location within the quarry at close-range with the intention of providing a high resolution reference point cloud. In the set-up used in this study, the range resolution of AVTIS2 was 0.75 m, much coarser than the 0.07 range resolution of the short-range TLS (Table \ref{Balmullo_Instrument_Summary}). Eight triangular trihedral corner reflectors were placed inside the quarry and were used as Ground Control Points (GCPs) (Fig. \ref{CC_Georeferencing}a) for georeferencing and AVTIS2 range calibration (see section \ref{AVTIS2_Processing}). We scanned across each reflector multiple times and averaged these to suppress noise, after which we fit a 2D Gaussian model (Fig. \ref{CC_Georeferencing}b) and extract its centre to represent the GCP angular position in the AVTIS2 radar frame of reference. The corresponding angular coordinates $\left(\theta,\phi\right)$ of each GCP are then converted to local cartesian coordinates $\left(\mathbf{G}_\text{Loc}\right)$. The geolocations of each instrument and the trihedral corner reflectors were determined using differential Global Positioning Systems (dGPS), where the maximum errors in $x$, $y$, and $z$ were $\Delta{x} = 1.2$ mm, $\Delta{y} = 1.5$ mm, and $\Delta{z} = 2.1$ mm, respectively.

% CORNER CUBE AND SCAN

\subsection{AVTIS2 Signal Processing} \label{AVTIS2_Processing}
\begin{figure}[!t]
	\centering
	\includegraphics[width=0.8\linewidth]{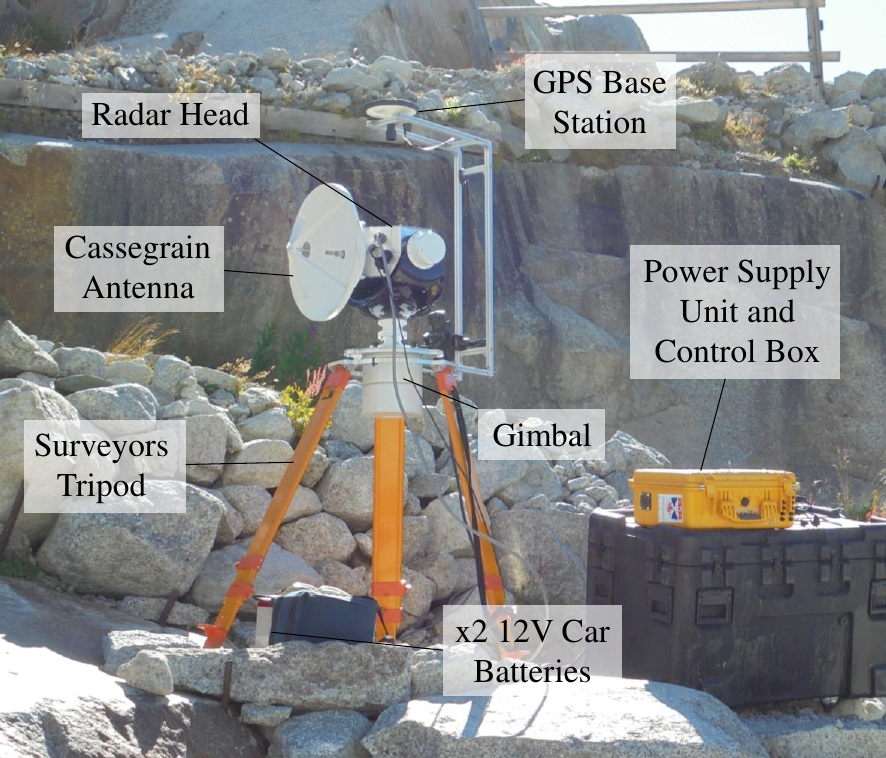}
	\caption{Annotated diagram of the AVTIS2 94 GHz radar system.}	
	\label{AVTIS2_Field_SetUp}
\end{figure}

% LEVEL-0 DATA ACQUISITION
AVTIS2 is a Frequency Modulated Continuous Wave (FMCW) radar that is deployed in the field on a surveyors' tripod, levelled, and a GPS bracket attached for geolocating the radar position (Fig. \ref{AVTIS2_Field_SetUp}). The radar is mounted onto a gimbal and is able to scan a full 360$\degree$ in azimuth, but is limited to between -16$\degree$ and 90$\degree$ in elevation, where 0$\degree$ is horizontal (i.e. the gimbal is levelled). The estimated radar pointing errors are primarily determined by the stated gimbal accuracy of 0.02$\degree$. At each site in this study, AVTIS2 scanned the same angular area twice to assess 3D point cloud precision. At Site 1 ($\sim$3.3 km), AVTIS2 scanned an angular area of 13$\degree$ in azimuth at 0.045$\degree$ increments and 2$\degree$ in elevation at 0.05$\degree$ increments, with scan times of $\sim$60-75 mins. Similarly at Site 2 ($\sim$1.4 km), AVTIS2 scanned an angular area of 25$\degree$ in azimuth at 0.045$\degree$ increments and 3.5$\degree$ in elevation at 0.05$\degree$ increments, with scan times of $\sim$60-75 mins. Along each LoS, AVTIS2 transmits a linearly ramped frequency signal with a fixed rate of change, called a chirp, from a circular cassegrain antenna and measures an Intermediate Frequency time series $\left(T_{\text{IF}}\right)$ which is then digitised using an Analog-to-Digital Converter (ADC). The raw digitised Level-0 data set represents $T_{\text{IF}}$ at each azimuth angle $\left(\theta\right)$ and elevation angle $\left(\phi\right)$ measured during an AVTIS2 scan and is time stamped with the control PC clock time to the start of the AVTIS2 scan. 

% AVTIS2 FIELD SET-UP

% LOS PROCESSING
A Fast Fourier Transform (FFT) is used to convert $T_{\text{IF}}$ to frequency $\left(f_{\text{IF}}\right)$ along each AVTIS2 LoS and a Blackman window applied to taper the edges and suppress spectral leakage \cite{Harris1978}. Range is then calculated using the FMCW equation:
\begin{equation}
R = \frac{f_{\text{IF}}cT_s}{2B}
\end{equation}
where $c$ is the speed of light ($3 \times 10^8$ m/s). The range resolution $\left(\Delta{R}\right)$ of the radar is given by:
\begin{equation}
\Delta{R} = \frac{c}{2B}
\end{equation}
from which the maximum unambiguous range $\left(R_{\text{max}}\right)$ can be calculated:
\begin{equation}
R_{\text{max}} = N_{\text{max}} \frac{c}{2B}
\end{equation}
where $N_{\text{max}}$ is number of samples in the single-sided FFT spectrum and is equal to 8,192 (i.e. there are 16,384 samples along each LoS). In this study, $B$ was set to 278 MHz (Site 1) and 299 MHz (Site 2), hence ${\Delta}R = 0.54$ m (Site 1) and ${\Delta}R = 0.5$ m (Site 2), respectively. The maximum unambiguous range of AVTIS2 was therefore $R_{\text{max}}=4,408$ m (Site 1) and $R_{\text{max}}=4,096$ m (Site 2), respectively. 

% DATA PROCESSING
Chirp non-linearity and range drift is corrected for using a range autofocussing technique which is based on the phase gradient algorithm \cite{Middleton2011}. This method calculates the AVTIS2 phase error by measuring the range to a stable trihedral corner reflector and comparing it to the vector distance determined by differential GPS (dGPS). AVTIS2 undertakes repeat measurements of this trihedral reflector before each scan of the terrain and warps the LoS signal in the time domain $\left(T_{\text{IF}}\right)$ by fitting an N\textsuperscript{th} order polynomial to the AVTIS2 phase error as a function of time. Corrupt spectra caused by radar trigger delays and issues with ADC time synchronisation are then removed, after which interference lines are suppressed, see details in \cite{Macfarlane2013}. An estimate of the AVTIS2 noise floor is then subtracted from the corrected AVTIS2 range profile of received power at each range bin to obtain a range profile in terms of Signal-to-Noise Ratio (SNR). The final product of AVTIS2 signal processing is therefore a 3D data cube of radar backscatter/SNR at each range bin $\left(R\right)$, azimuth angle $\left(\theta\right)$, and elevation angle $\left(\phi\right)$ measured during an AVTIS2 scan. 

% NOISE FLOOR CORRECTION
\begin{figure}[!t]
	\centering
	\includegraphics[width=0.9\linewidth]{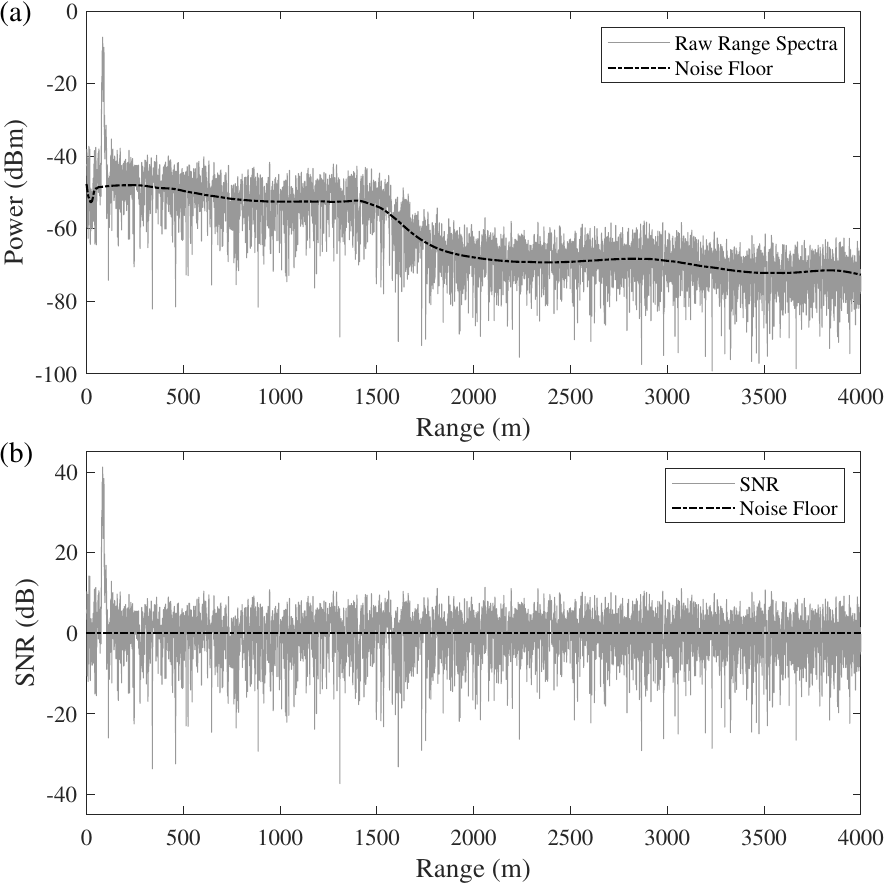}
	\caption{(a) Raw range profile overlaid with the average AVTIS2 noise floor. These are differenced to create the SNR profile shown in (b). The average AVTIS2 noise floor is estimated for each AVTIS2 scan by noncoherently averaging measurements of the sky or, when a sky estimate is not available, averaging all range profiles at a single elevation angle and manually removing terrain returns.}
	\label{NoiseFloor_Correction}
\end{figure}

\subsection{Point Cloud Comparison}
\begin{figure}[!t]
	\centering
	\includegraphics[width=\linewidth]{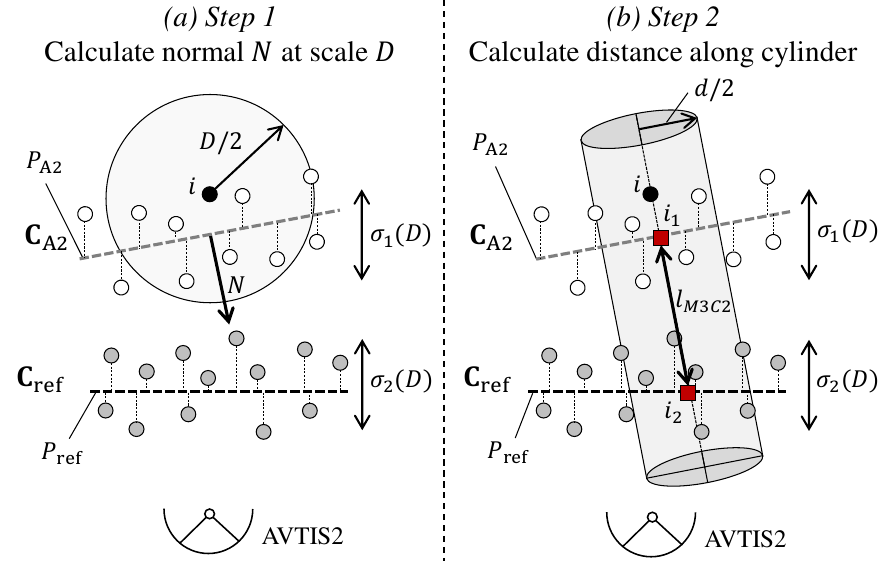}
	\caption{Overview of the M$3$C$2$ method. (a) Step 1 takes a point $i$ in the point cloud $\mathbf{C}_{\text{A2}}$, extracts neighbouring points within a search sphere of diameter $D/2$, and calculates the local surface normal $N$ relative to the surface $P_{\text{A2}}$. (b) In Step 2, this surface normal is used to construct a search cylinder, where points from both $\mathbf{C}_{\text{A2}}$ and $\mathbf{C}_{\text{ref}}$ are extracted and the local mean C2C distance calculated to estimate the M3C2 distance $\left(l_{\text{M3C2}}\right)$.}	
	\label{M3C2_Method}
\end{figure}

% M3C2 METHOD: STEP 1
The distance between the AVTIS2 and short-range TLS point clouds extracted over Balmullo Quarry was used to quantify the uncertainty of AVTIS2 3D point clouds. Here, we use the Multiscale Model to Model Cloud Comparison (M3C2) method developed by Lague et al. \cite{Lague2013} (Fig. \ref{M3C2_Method}), which calculates the local mean C2C distance between two point clouds, one under test $\left(\mathbf{C}_{\text{com}}\right)$ and a reference $\left(\mathbf{C}_{\text{ref}}\right)$. In step 1, the M3C2 method defines a set of core points in $\mathbf{C}_{\text{com}}$ to compute the point cloud difference. Because the AVTIS2 point cloud density is low, we define each point in the AVTIS2 point cloud $\left(\mathbf{C}_{\text{A2}}\right)$ as a core point. The points surrounding each point in $\mathbf{C}_{\text{A2}}$ $\left(\mathbf{c}_{\text{nn}}\right)$ are extracted by defining a search sphere with radius $D/2$ (Fig. \ref{M3C2_Method}). The value of $D/2$ is estimated from the size of the beam footprint:
\begin{align}
\label{Beam_Footprint_Equation}
D/2=R\tan{\left(\frac{\theta_2}{2}\right)}
\end{align}
where $R$ is range and $\theta_2$ is the two-way radar beamwidth in azimuth $\left(0.33\degree\right)$. The M3C2 algorithm uses a single value of $D/2$ and thus is not adaptable to points at different ranges. Therefore, $R$ is set to the average distance of the AVTIS2 radar at each site i.e. $R=3.3$ km (Site 1) and $R=1.4$ km (Site 2), giving values of $D/2=10$ m (Site 1) and $D/2=4$ m (Site 2), respectively. A local plane $\left(P_{\text{A2}}\right)$ is fitted to these neighbouring points $\left(\mathbf{c}_{\text{nn}}\right)$ and used to calculate the local surface normal of the 3D surface $\left(N\right)$, where $N$ is oriented positively towards the radar (Fig. \ref{M3C2_Method}a).

% M3C2 METHOD: STEP 2
In step 2, the local surface normal is used to define the axis along which the point cloud comparison at each point in $\mathbf{C}_{\text{A2}}$ is computed. Along this axis, a search cylinder of diameter $d/2$ is defined and is taken to be the same as $D/2$ (Fig. \ref{M3C2_Method}b). All points lying within this search cylinder from both point clouds are extracted and the average position of each computed ($i_1$ and $i_2$). The Euclidean distance between $i_1$ and $i_2$ is then calculated to estimate  $l_{\text{M3C2}}$ i.e. the M3C2 point cloud distance between each point in $\mathbf{C}_{\text{A2}}$ with respect to $\mathbf{C}_{\text{ref}}$. The Level of Detection (LoD) at the 95\textsuperscript{th} confidence interval is used to estimate the associated uncertainty with the M3C2 distance calculation:
\begin{align}
\text{LoD}_{95\%} = \pm1.96 \left( \sqrt{ \dfrac{\sigma_1\left(d\right)^2}{n_1} + \dfrac{\sigma_2\left(d\right)^2}{n_2} } \right)
\end{align}
where $\sigma_1\left(d\right)$ and $\sigma_2\left(d\right)$ are the local point cloud roughnesses in the search area $D/2$ relative to their respective local planes $\left(P_{\text{A2}}\text{ and }P_{\text{ref}}\right)$ (Fig. \ref{M3C2_Method}), whilst $n_1$ and $n_2$ are the corresponding number of points in $\mathbf{c}_{\text{nn}}$. The M3C2 method accounts for differences in point cloud densities by averaging the position of each point cloud within a cylinder, hence reducing the impact of local surface roughness. Further, orienting the cylinder in the direction of the local surface normal ensures that differences between two point clouds occur perpendicular to the terrain rather than in any random direction.

\subsection{3D Point Cloud Uncertainty}
% UNCERTAINTY
The total uncertainty of the AVTIS2 point cloud $\left(\sigma_{\text{M3C2}}\right)$ is taken to be the standard deviation of $l_{\text{M3C2}}$ whilst the average of $l_{\text{M3C2}}$ $\left(\overline{l}\right)$ indicates the existence of systematic errors associated with the AVTIS2 measurements. Here, the M3C2 point cloud comparison is computed in the direction of the radar origin, called $-{\left(0,0,0\right)}$ in the Cloud Compare software \cite{Lague2013}, hence positive values indicate the range to terrain is underestimated and negative values indicate the range to terrain is overestimated. Several sources of uncertainty contribute to $\sigma_{\text{M3C2}}$ which can be combined using uncertainty propagation:
\begin{align}
\label{4_M3C2_Uncertainty}
\sigma^2_{\text{M3C2}} = \sigma^2_{\text{A2}} + \sigma^2_{\text{ref}} + \sigma^2_{l}
\end{align}
Here, $\sigma_{\text{A2}}$ is the AVTIS2 positional uncertainty, $\sigma_{\text{ref}}$ is the stated positional uncertainty of the reference point cloud (i.e. derived from the short-range TLS) which is $6$ mm \cite{Leica2011}, and $\sigma_{l}$ is the M3C2 distance uncertainty given by $\text{LoD}_{95\%}$ at each point in $\mathbf{C}_{\text{A2}}$. The combined uncertainty of $\sigma_{l}$ and $\sigma_{\text{ref}}$ represents the total uncertainty due to the M3C2 distance computation and is independent from AVTIS2 radar uncertainty $\left(\sigma_{\text{A2}}\right)$:
\begin{align}
{\Delta{E}}^2 = \sigma^2_{\text{ref}} + \sigma^2_{l}
\end{align}
A systematic error within the AVTIS2 point cloud $\left(\mathbf{C}_{\text{A2}}\right)$ can only be detected when $\overline{l} > \Delta{E}$. If $\sigma_{\text{M3C2}} < \Delta{E}$, then the M3C2 uncertainty dominates and the AVTIS2 uncertainty cannot be measured. Overall, the following rules are applied for calculating the AVTIS2 positional uncertainty $\left(\sigma_{\text{A2}}\right)$:
\begin{align}
\sigma_{\text{A2}} \approx
\begin{cases}
\sigma_{\text{M3C2}} ,& \text{if } \sigma_{\text{M3C2}} \geq \sigma_{l} \text{ and } \sigma_{\text{M3C2}} \geq \sigma_{\text{ref}} \\
\sigma_{l} ,& \text{if } \sigma_{l} \geq \sigma_{\text{M3C2}} \geq \sigma_{\text{ref}} \\
\sigma_{\text{ref}} ,& \text{if } \sigma_{\text{ref}} \geq \sigma_{\text{M3C2}} \geq \sigma_{l} \\
\Delta{E} ,& \text{if } \sigma_{l} \geq \sigma_{\text{M3C2}} \text{ and } \sigma_{\text{ref}} \geq \sigma_{\text{M3C2}}
\end{cases}
\end{align}

\section{Results} \label{Section_Results}
\subsection{Algorithm Comparison: Original vs \textit{PCFilt-94}}
% INTRODUCTION
In this section, the performance of the new \textit{PCFilt-94} algorithm is evaluated. To do this, we compare the accuracy of the AVTIS2 point clouds extracted using the new \textit{PCFilt-94} algorithm and the old range to maximum SNR technique. Comparisons are made by georeferencing each AVTIS2 point cloud on each day using the maximum number of available GCPs: 7 at Site 1 and 8 at Site 2. A summary of the comparison is shown in Table \ref{Waveform_Averaging_Stats}.

% SUMMARY STATISTICS
Overall, the \textit{PCFilt-94} algorithm improves the accuracy of the point cloud at both short and long range (Table \ref{Waveform_Averaging_Stats}). At Site 1 (3.3 km), applying the \textit{PCFilt-94} algorithm reduces $\sigma_{\text{A2}}$ from $\pm$3.17 m to $\pm$2.75 m, leading to a $\pm$0.42 m (13\%) reduction in point cloud uncertainty. Similarly at Site 2 (1.4 km), the application of the \textit{PCFilt-94} algorithm reduces $\sigma_{\text{A2}}$ from $\pm$1.85 m to $\pm$1.48 m, which is a reduction of $\pm$0.37 m (20\%) in point cloud uncertainty. $\sigma_{\text{A2}}$ also increases with range, hence point cloud uncertainty increases with distance from the radar and is an inherent limitation of real-beam radars. This effect is independent of whether or not the \textit{PCFilt-94} algorithm is applied given that the reduction in $\sigma_{\text{A2}}$ is approximately $\pm$0.4 m in both cases. Applying the \textit{PCFilt-94} algorithm also reduces the M3C2 error $\left(\Delta{E}\right)$. This makes the error component $\overline{l}$ larger than $\Delta{E}$ when the \textit{PCFilt-94} algorithm is applied and reveals the existence of a small systematic point cloud offset of -0.38 m at Site 1 (3.3 km) and 0.35 m at Site 1 (1.4 km). However, the values of $\overline{l}$ and $\Delta{E}$ when the \textit{PCFilt-94} algorithm is applied are both below the AVTIS2 range resolution $\left(\Delta{R}\right)$ which was set to 0.54 m at Site 1 (3.3 km) and 0.5 m at Site 2 (1.4 km), hence are not considered significant.

% WAVEFORM AVERAGING STATISTICS TABLE
\begin{table}[!t]
\caption{Key statistics relating to the performance of the waveform averaging technique.}
\label{Waveform_Averaging_Stats}
\centering
\begin{tabular}{|c|c|c|c|c|}
\hline
\textbf{Site} & \textbf{Algorithm} & \textbf{$\sigma_{\text{A2}}$ (m)} & \textbf{$\bar{l}$ (m)} & \textbf{$\Delta{E}$ (m)}  \\
\hline
Site 1 (3.3 km) & Original & $\pm$3.17 & -0.42 & $\pm$1.02 \\
\hline
Site 1 (3.3 km) & \textit{PCFilt-94} & $\pm$2.75 & -0.38 & $\pm$0.32 \\
\hline
Site 2 (1.4 km) & Original & $\pm$1.85 & 0.13 & $\pm$0.70 \\
\hline
Site 2 (1.4 km) & \textit{PCFilt-94} & $\pm$1.48 & 0.35 & $\pm$0.18 \\
\hline
\end{tabular}
\end{table}

\begin{figure}[!t]
	\centering
	\includegraphics[width=\linewidth]{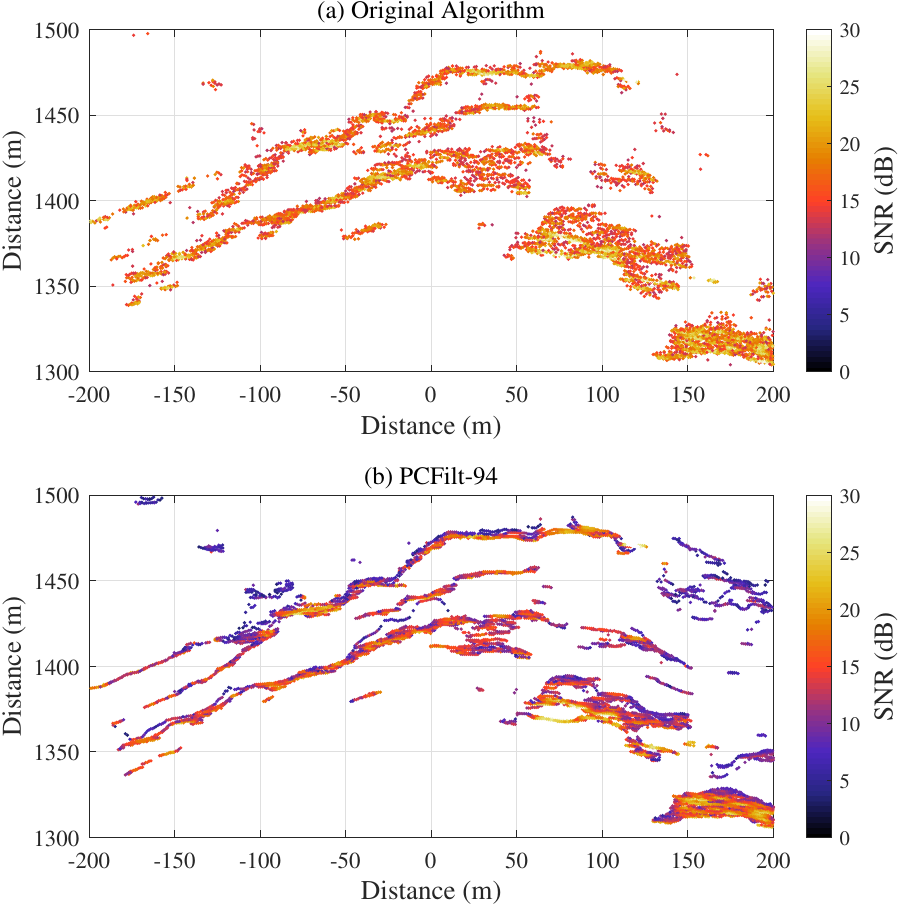}
	\caption{Impact of applying the waveform averaging technique during the point cloud extraction process. (a) Original algorithm, and (b) point cloud extracted from \textit{PCFilt-94}.}	
	\label{WaveformAveraged_vs_Raw_PointCloud}
\end{figure}

% VISUAL COMPARISON OF WAVEFORM AVERAGING
The point cloud extracted without waveform averaging is more variable compared to when the waveform averaging is applied as shown in Fig. \ref{WaveformAveraged_vs_Raw_PointCloud} as a result of radar speckle. The larger uncertainty of the raw point cloud is due to the greater local spatial variability of the points representing terrain, suggesting that fluctuations in the raw waveform data can lead to less accurate retrievals of terrain using the range to maximum SNR algorithm. A significant consequence of the more variable point cloud that is not reflected in the uncertainty statistics is that the Voronoi-based spatial outlier filtering identifies a greater proportion of points in the raw point cloud as outliers. Applying the point cloud filtering processing to both point clouds at Site 1 (3.3 km) led to the extraction of 8,019 (waveform averaging) and 6,107 (no waveform averaging) points, respectively, which is a reduction of 2,002 points when no averaging is applied. Similarly at Site 2 (1.4 km), a total of 25,774 (waveform averaging) and 10,291 (no waveform averaging) points were extracted, respectively, which is a reduction of 15,483 points when no averaging is applied. Therefore, a critical performance enhancement of the waveform averaging technique is the improved stability of the resultant point cloud and its lower sensitivity to random fluctuations along a waveform when extracting the range to terrain.

% WAVEFORM AVERAGING EXAMPLE

\begin{figure}[!t]
	\centering
	\includegraphics[width=\linewidth]{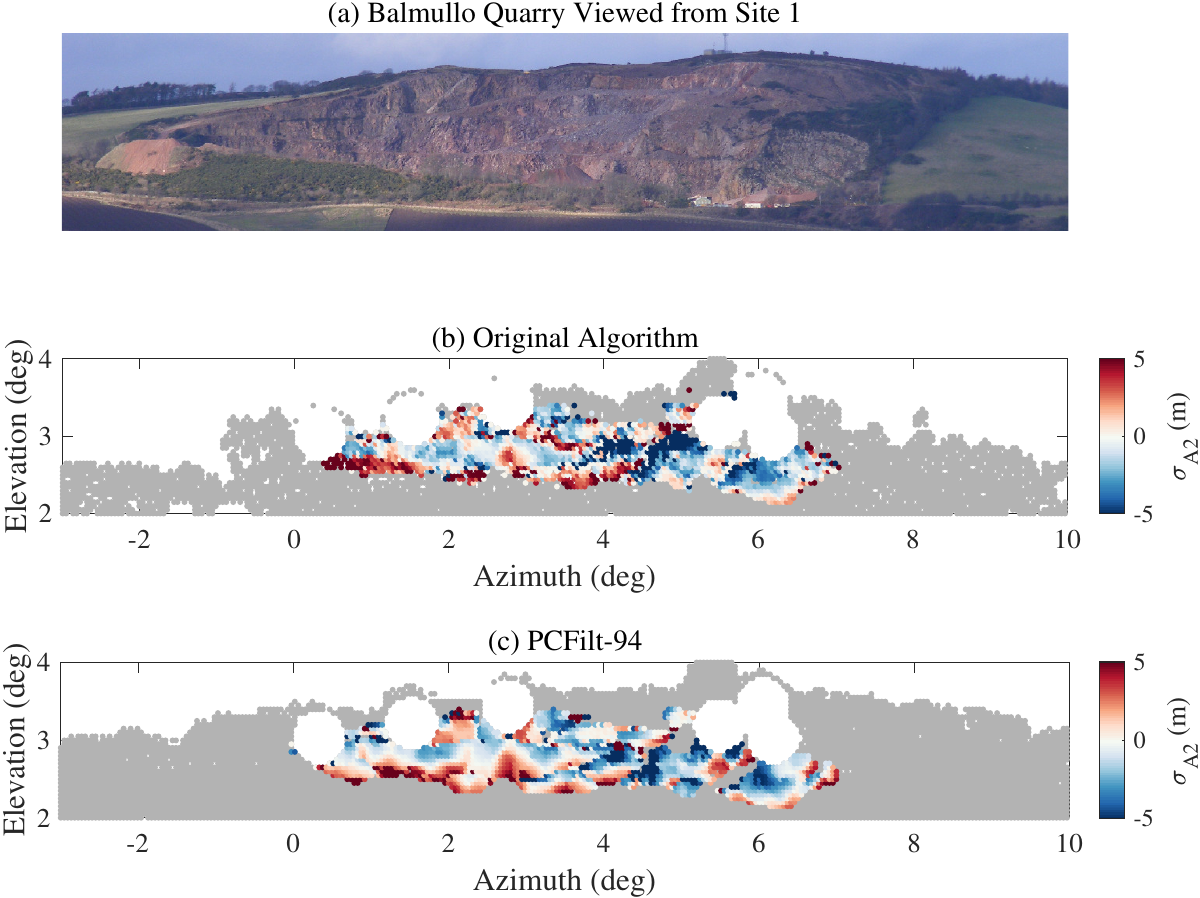}
	\caption{Performance of the waveform averaging technique on long-range mapping of Balmullo Quarry at Site 1. (a) Balmullo as viewed from Site 1, (b) $\sigma_{\text{A2}}$ when no waveform averaging is applied, and (c) $\sigma_{\text{A2}}$ when waveform averaging is applied. In plots (b) and (c), the full point cloud is plotted in grey whilst the points overlapping with the short-range TLS are plotted as a function of $\sigma_{\text{A2}}$. Positive (red) values indicate that points are offset closer to the radar, whilst negative (blue) values indicate that points are offset at a distance further away from the reference point cloud relative to the radar. Note the AVTIS2 spot size at 3.3 km is $\sim$20.2 m. Note, the holes in the point cloud represent the locations of trihedral reflectors which were removed during pre-processing.}	
	\label{Waveform_Averaging_LongRange}
\end{figure}

% SITE 1 WAVEFORM AVERAGING SPATIAL PERFORMANCE
The spatial variability in $\sigma_{\text{A2}}$ at both sites reveals the impact of waveform averaging on point cloud uncertainty. At Site 1 (3.3 km), the raw point cloud variability is smoothed when the waveform averaging is applied (Fig. \ref{Waveform_Averaging_LongRange}c), thus removing small scale variability in point cloud positional uncertainty. Locations where $\sigma_{\text{A2}}$ is large (i.e towards the bottom of Figs. \ref{Waveform_Averaging_LongRange}b and \ref{Waveform_Averaging_LongRange}c) are coincident with regions of vegetation, suggesting that volume backscatter from such targets \cite{Ulaby1990} is large and inhibits signal penetration through to the quarry face. Across the non-vegetated quarry face, $\sigma_{\text{A2}}$ is largely negative and thus overestimates the range to terrain at long range. Because the beam spot size at 3.3 km is $\sim$20 m, the radar illuminates a large region of the quarry at long-range. Therefore, any single scattering object with a large Radar Cross Section $\left(\sigma\right)$ within a beam footprint (e.g. the apex of a quarry cliff) can dominate the terrain returns at multiple elevation angles and obstruct the returns of the underlying topographic signal. The result is a systematic offset of the point cloud locally and causes a gradual change in $\sigma_{\text{A2}}$ as is observed in Figs. \ref{Waveform_Averaging_LongRange}b and \ref{Waveform_Averaging_LongRange}c and hence cannot be suppressed using waveform averaging.

% SITE 1 WAVEFORM AVERAGING PERFORMANCE Fig.

\begin{figure}[!t]
	\centering
	\includegraphics[width=\linewidth]{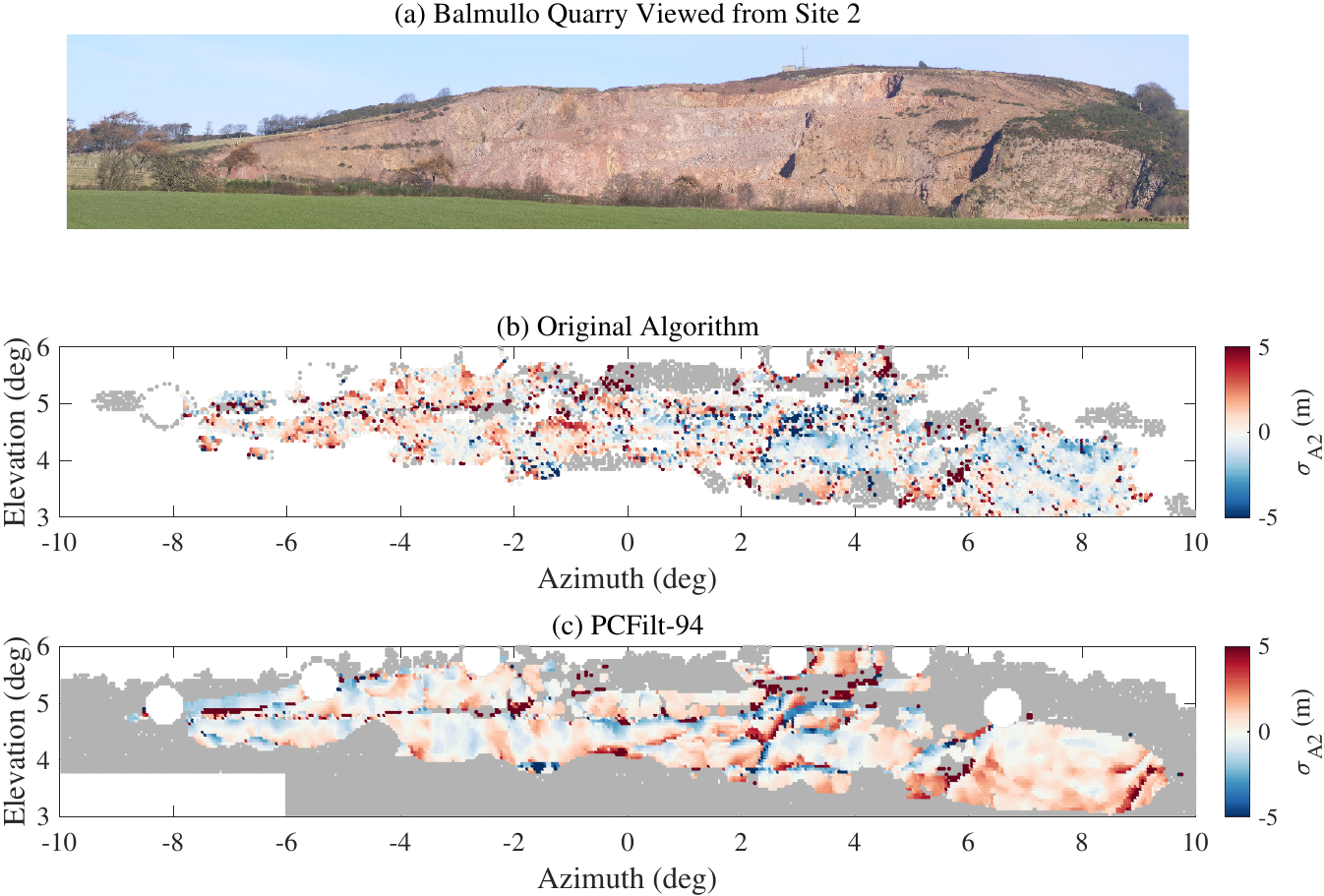}
	\caption{Performance of the waveform averaging technique on short-range mapping of Balmullo Quarry at Site 2. (a) Balmullo as viewed from Site 2, (b) $\sigma_{\text{A2}}$ when no waveform averaging is applied, and (c) $\sigma_{\text{A2}}$ when waveform averaging is applied. In plots (b) and (c), the full point cloud is plotted in grey whilst the points overlapping with the short-range TLS are plotted as a function of $\sigma_{\text{A2}}$. Positive (red) values indicate that points are offset closer to the radar, whilst negative (blue) values indicate that points are offset at a distance further away from the reference point cloud relative to the radar. Note the AVTIS2 spot size at 1.4 km is $\sim$8.6 m.}	
	\label{Waveform_Averaging_ShortRange}
\end{figure}

% SITE 2 WAVEFORM AVERAGING SPATIAL PERFORMANCE
At shorter range (Site 2 at 1.4 km), the waveform averaging method clearly suppresses noise (Fig. \ref{Waveform_Averaging_ShortRange}). Across the vegetation-free components of the quarry surface, $\sigma_{\text{A2}}$ undulates between positive and negative values as a result of the smoothing effect of the waveform averaging which induces a minor offset in the signal echo position within the resulting averaged waveform. This effect is generally smaller than at Site 1. Regions of locally higher $\sigma_{\text{A2}}$ correlate with the locations of discontinuities across the quarry where the incidence angle to the terrain is small. Along these edges, $\sigma_{\text{A2}}$ is more negative at the corners of the surface nearest the radar, but more positive at the edges of the surface furthest from the radar. This indicates that the large radar footprint along these discontinuities smooths the signal between the two illuminated surfaces and calculates the range to terrain as an average of both. 

% SITE 2 WAVEFORM AVERAGING PERFORMANCE Fig.

\subsection{Single vs Multiple Point Extraction} \label{Single_Multiple_Section}
% SINGLE VS MULTIPLE POINT PROCESSING
The relative performances of the single and multiple point extraction methodologies is summarised in Table \ref{Single_Multiple_Stats}. Furthermore, Table \ref{Total_Number_of_Points} summarises the total number of points in the respective point clouds extracted in this study. In general, the multiple-point processing methodology increases the uncertainty $\left(\sigma_{\text{A2}}\right)$ of the point cloud at both sites. At Site 1 (3.3 km), $\sigma_{\text{A2}}$ is $\pm$2.75 m for the single point methodology and $\pm$3.12 m for the multiple point methodology, respectively, resulting in a $\pm$0.37 m (27\%) increase in point cloud uncertainty when extracting additional points (Table \ref{Single_Multiple_Stats}). At Site 2 (1.4 km), point cloud uncertainty increases from $\pm$1.48 m (single point) to $\pm$2.05 m (multiple point), leading to an increase in point cloud uncertainty of $\pm$0.57 m (39\%) when applying the multiple point methodology at shorter range (Table \ref{Single_Multiple_Stats}). Therefore, whilst the uncertainty of point clouds extracted using the multiple-point processing methodology increases with range, it more significantly increases point cloud uncertainty at short range. A minor systematic offset characterised by $\overline{l}>\Delta{E}$ is observed at both study locations, although not for the multiple point extraction cloud at Site 1 (3.3 km). This offset is considered negligible given that it is below both the total point cloud uncertainty $\left(\sigma_{\text{A2}}\right)$ and AVTIS2 range resolution (0.54 m at Site 1 (3.3 km) and 0.5 m at Site 2 (3.3 km)) in all cases.

% SINGLE VS MULTIPLE STATISTICS
\begin{table*}[!t]
\caption{Summary statistics of the single and multiple point extraction methodology performance. The `Single-Multiple Difference' represents the statistics for the additional points extracted using the multiple point methodology.}
\label{Single_Multiple_Stats}
\centering
\begin{tabular}{|c|c|c|c|c|c|}
\hline
\textbf{Site} & \textbf{Point Cloud} & \textbf{Number of Points} & \textbf{$\sigma_{\text{A2}}$ (m)} & \textbf{$\overline{l}$ (m)} & \textbf{$\Delta{E}$ (m)}  \\
\hline
Site 1 (3.3 km) & Single & 8,019 & $\pm$2.75 & -0.38 & $\pm$0.32 \\
\hline
Site 1 (3.3 km) & Multiple & 14,671 & $\pm$3.12 & 0.15 & $\pm$0.31 \\
\hline
Site 1 (3.3 km) & Single-Multiple Difference & 6,652 & $\pm$4.00 & 1.51 & $\pm$0.36 \\
\hline
Site 2 (1.4 km) & Single & 25,774 & $\pm$1.48 & 0.35 & $\pm$0.18 \\
\hline
Site 2 (1.4 km) & Multiple & 41,364 & $\pm$2.05 & 0.45 & $\pm$0.21 \\
\hline
Site 1 (1.4 km) & Single-Multiple Difference & 15,590 & $\pm$3.26 & 0.84 & $\pm$0.35 \\
\hline
\end{tabular}
\end{table*}

% TABLE: TOTAL NUMBER OF POINTS FOR EACH INSTRUMENT.
\begin{table}[!t]
\caption{Total number of points contained within each AVTIS2 point cloud and the percentage overlap with the short-range TLS point cloud which contained 25,414,521 points.}
\label{Total_Number_of_Points}
\centering
\begin{tabular}{|c|c|c|c|}
\hline
\textbf{Sensor} & \textbf{Site} & \textbf{Points} & \textbf{Overlap} \\
\hline
AVTIS2 (Single) & Site 1 (3.3 km) & 8,019 & 2,025 (25.25\%) \\
\hline
AVTIS2 (Multiple) & Site 1 (3.3 km) & 14,671 & 2,846 (19.40\%) \\
\hline
AVTIS2 (Single) & Site 2 (1.4 km) & 25,774 & 11,772 (45.67\%) \\
\hline
AVTIS2 (Multiple) & Site 2 (1.4 km) & 41,364 & 15,806 (38.21\%) \\
\hline
\end{tabular}
\end{table}

\begin{figure*}[!t]
	\centering
	\includegraphics[width=0.75\linewidth]{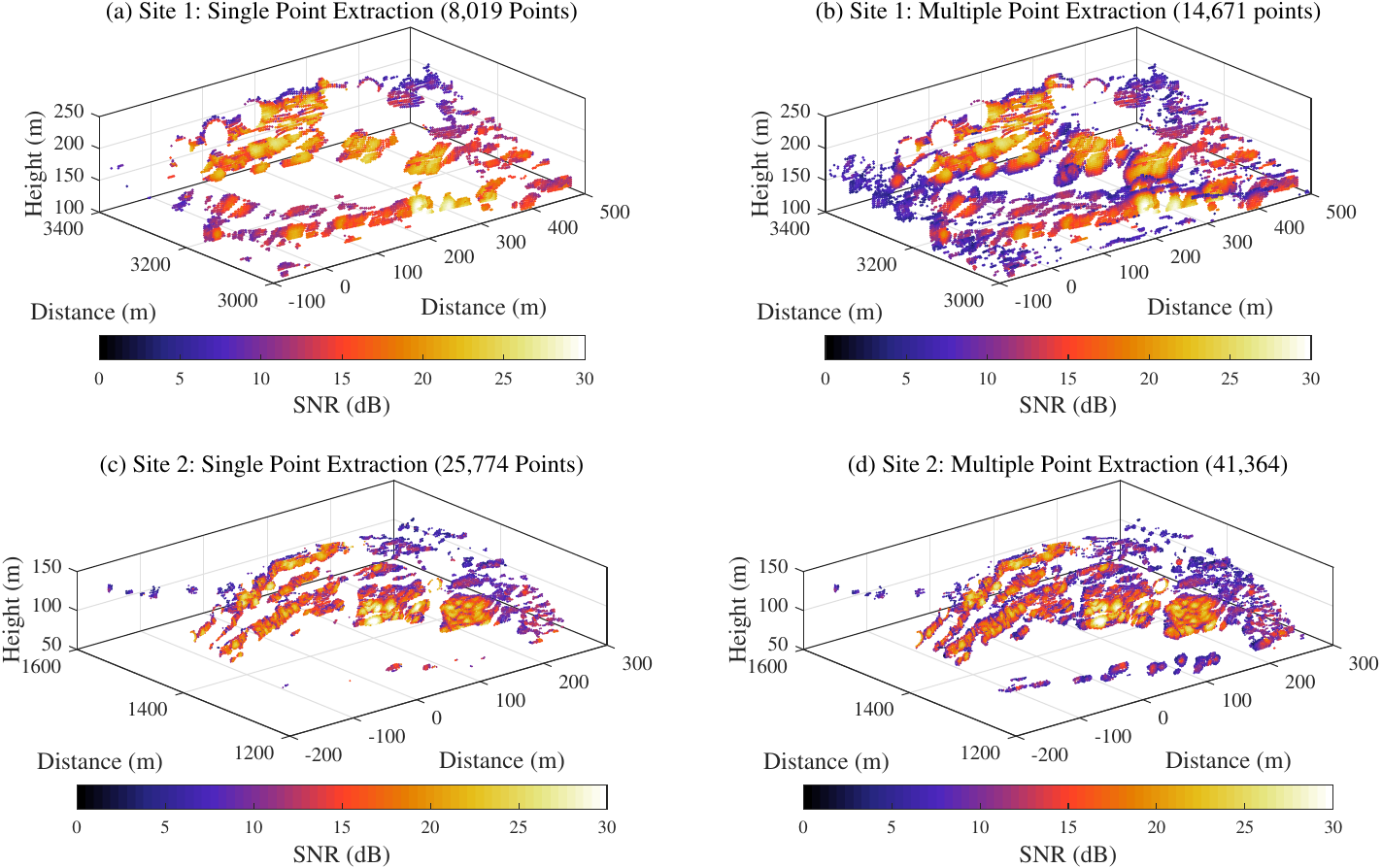}
	\caption{The four points clouds extracted during this study. At Site 1, (a) the single point and (b) multiple point extraction methodologies are shown. Similarly from Site 2, (c) the single point and (d) multiple point extraction methodologies are shown.}	
	\label{Single_Multiple_PointCloud}
\end{figure*}

% POINT CLOUD COMPARISON
Examples of point clouds extracted using both methodologies are shown in Fig. \ref{Single_Multiple_PointCloud}. The number of points extracted using the multiple point methodology increased by 83\% at Site 1 (3.3 km) and by 60\% at Site 2 (1.4 km). The uncertainty contribution from these additional points ('Single-Multiple Difference' (SMD) in Table \ref{Single_Multiple_Stats}) is $\pm$4.00 m at Site 1, which is $\pm$0.88 m (28\%) larger than the multiple point cloud, and $\pm$3.26 m at Site 2 (1.4 km), which is $\pm$1.21 m (59\%) larger than the multiple point cloud. Also, there is a significant systematic offset $(\overline{l})$ between the SMD points and the terrain. At Site 1 (3.3 km) the offset is 1.51 m and at Site 2 (1.4 km) the offset is 0.84 m, hence in both cases the additional points underestimate the range to terrain. The larger errors and uncertainty of the additional points extracted in the multiple point methodology increases the total uncertainty of the point cloud and degrades the point cloud accuracy. This is likely due to the lower SNR of the additional points. At Site 1 (3.3 km), the mean SNR of the total point cloud was 14.61 dB (single) and 8.93 dB (multiple), whilst at Site 2 (1.4) the mean SNR of the total point cloud was 17.54 dB (single) and 12.87 dB (multiple). Therefore, the multiple point methodology extracts additional points with lower SNR compared to the bulk terrain returns which have a lower probability of being detected in subsequent scans of the terrain and may be missed if radar noise power or atmospheric attenuation increases. Overall, the increased uncertainty of the multiple point extraction methodology is generally small and in circumstances where higher point cloud densities are desired (e.g. for change detection), the multiple point methodology is sufficiently accurate.

% SINGLE AND MULTIPLE POINT CLOUD

\begin{figure}[!t]
	\centering
	\includegraphics[width=\linewidth]{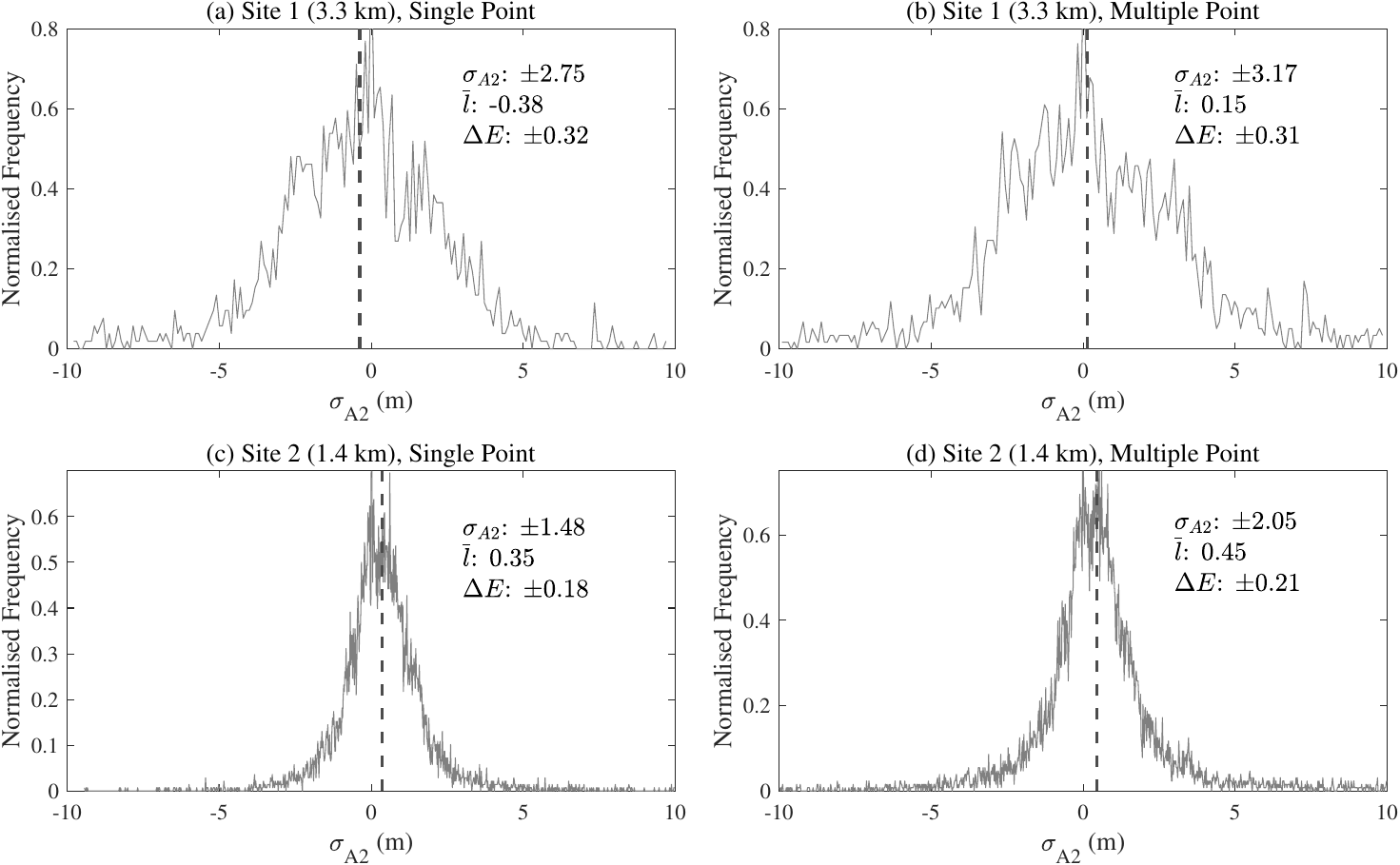}
	\caption{The four points clouds extracted during this study. At Site 1, (a) the single point and (b) multiple point extraction methodologies are shown. Similarly from Site 2, (c) the single point and (d) multiple point extraction methodologies are shown.}	
	\label{Single_Multiple_Point_Processing}
\end{figure}

% HISTOGRAM CHARACTERISTICS
Histograms of $\sigma_{\text{A2}}$ for the single and multiple point methodologies at each site are shown in Fig. \ref{Single_Multiple_Point_Processing}. Each histogram is narrow and have long tails, particularly at Site 2 (1.4 km) (Figs. \ref{Single_Multiple_Point_Processing}c and \ref{Single_Multiple_Point_Processing}d), which is broadly characteristic of a Lorentzian distribution. This suggests that there is a higher probability of point cloud errors exceeding $\pm\sigma_{\text{A2}}$. This is consistent with the results from the previous section which showed that discontinuities across the quarry surface cause sudden changes in point cloud uncertainty. Such complex topographic environments may be present across other natural surfaces and so the Lorentzian distribution may also be characteristic of AVTIS2 point cloud uncertainties at other locations, but additional experiments are required to confirm this. At Site 1 (3.3 km), the spread of the probability distribution is broader which reflects the larger uncertainty of longer range point clouds (Figs. \ref{Single_Multiple_Point_Processing}a and \ref{Single_Multiple_Point_Processing}b). Further, the point cloud extracted using the multiple point methodology at Site 1 (3.3 km) is skewed to negative values of $\sigma_{\text{A2}}$ (Fig. \ref{Single_Multiple_Point_Processing}b) as a result of the systematic offset in the additional points (SMD) introduced into this point cloud. 

\subsection{Point Cloud Repeatability (Precision)} \label{PointCloud_Precision}
\begin{figure}[!t]
	\centering
	\includegraphics[width=\linewidth]{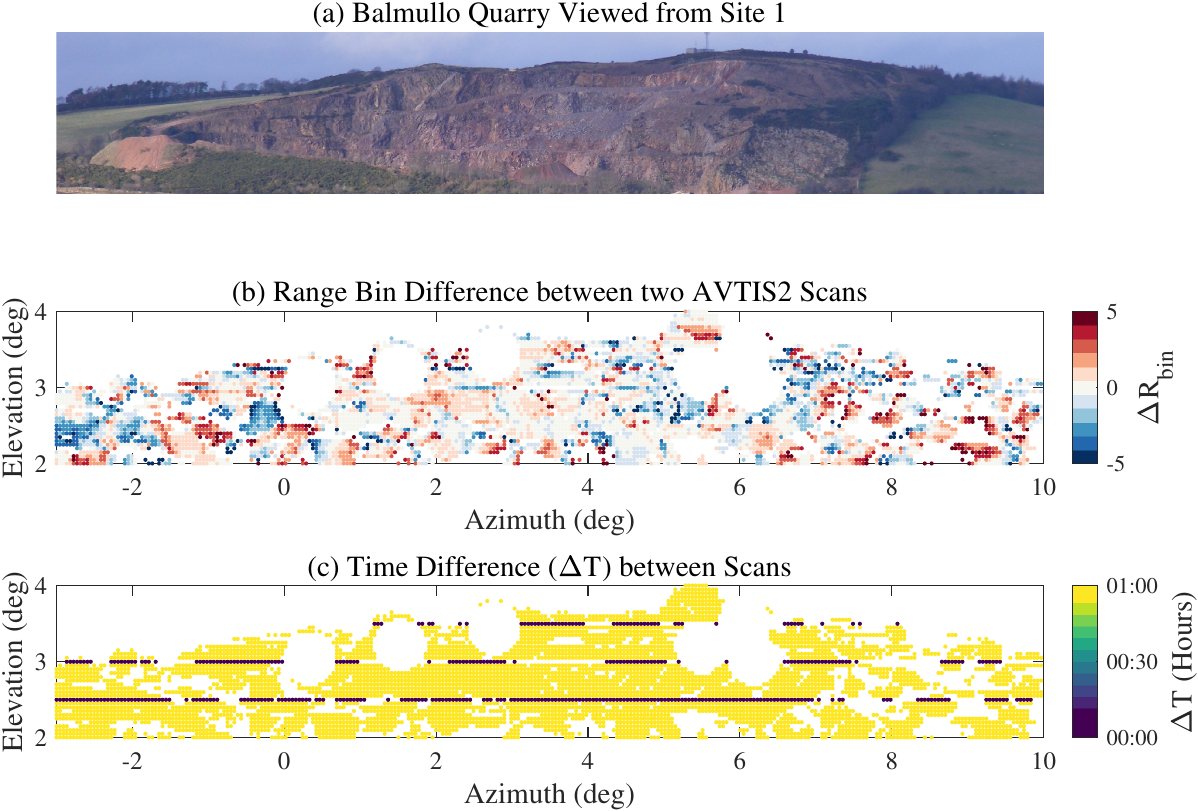}
	\caption{(a) Balmullo Quarry viewed from Site 1, (b) range bin difference $\left(\Delta{R_{\text{bin}}}\right)$ between consecutive scans of the terrain at each angular position across the AVTIS2 Field of View (FoV), and (c) the time difference between measurements at each angular position.}	
	\label{AVTIS2_Repeatability_LongRange}
\end{figure}

\begin{figure}[!t]
	\centering
	\includegraphics[width=\linewidth]{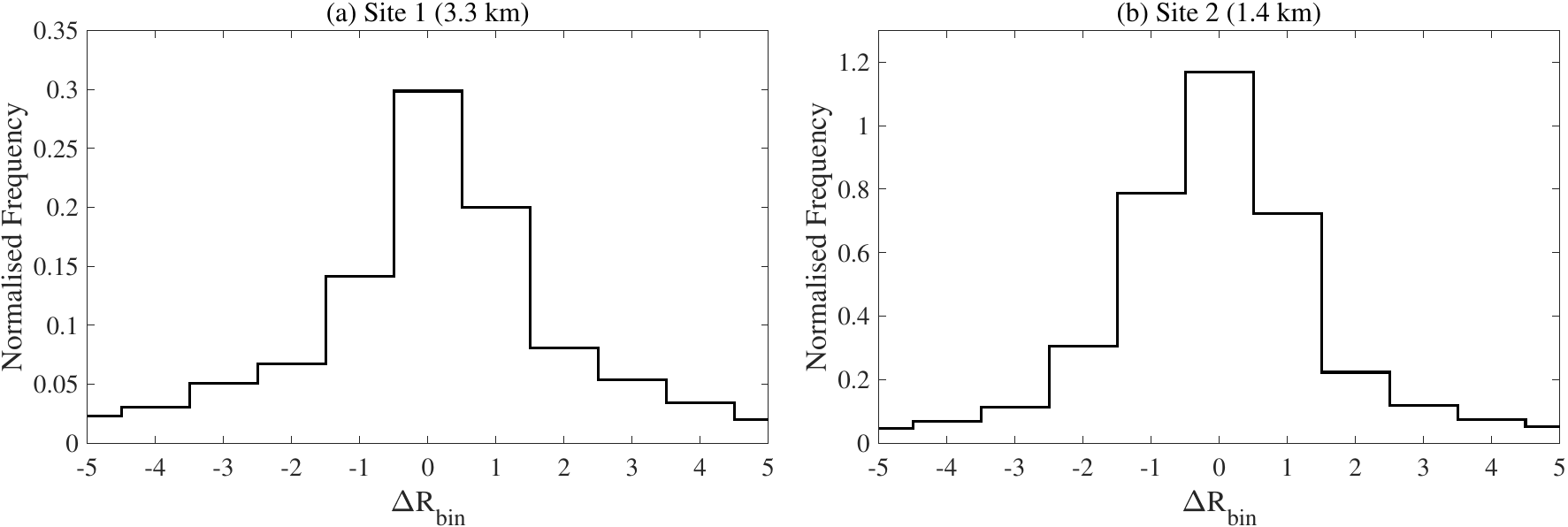}
	\caption{Histograms showing the range bin difference $\left(\Delta{R_{\text{bin}}}\right)$ between consecutive AVTIS2 scans of the Balmullo Quarry topography at (a) Site 1 (3.3 km) and (b) Site 2 (1.4 km).}	
	\label{AVTIS2_Repeatability_Histograms}
\end{figure}

% POINT CLOUD REPEATBILITY: LONG RANGE
The repeatability of an AVTIS2 point cloud extracted using the \textit{PCFilt-94} algorithm is evaluated by differencing the range bin position $\left(\Delta{R_{\text{bin}}}\right)$ of terrain along each Line of Sight (LoS) in two consecutive scans of Balmullo Quarry. At Site 1 (3.3 km), $\Delta{R_{\text{bin}}}$ can be as large as $\pm$5 bins (Fig. \ref{AVTIS2_Repeatability_LongRange}a), which is equivalent to $\pm$2.65 m and thus close to the uncertainty of long-range AVTIS2 point clouds calculated in section \ref{Single_Multiple_Section}. Locations of high $\Delta{R_{\text{bin}}}$ (red) are generally coincident with regions of vegetation and discontinuities in terrain, hence complex topography can significantly reduce the ranging precision of AVTIS2 measurements. The histogram of $\Delta{R_{\text{bin}}}$ at Site 1 (3.3 km) indicates that the range to terrain over most of the quarry varies by $\pm$1 range bin (Fig. \ref{AVTIS2_Repeatability_Histograms}a). However, a more conservative estimate based on the standard deviation of $\Delta{R_{\text{bin}}}$ gives a value of $\pm$2 range bins. This is an important result when considering using AVTIS2 point clouds for change detection. At Site 1 (3.3 km), the point cloud variability appears to be controlled by the complex topography of the quarry and radar hardware limitations as opposed to atmospheric distortion of the signal given that the time difference between consecutive AVTIS2 measurements was less than 1 hour (Fig. \ref{AVTIS2_Repeatability_LongRange}c).

% POINT CLOUD REPEATBILITY PLOT: LONG RANGE

% POINT CLOUD REPEATBILITY PLOT: SHORT RANGE

\begin{figure}[!t]
	\centering
	\includegraphics[width=\linewidth]{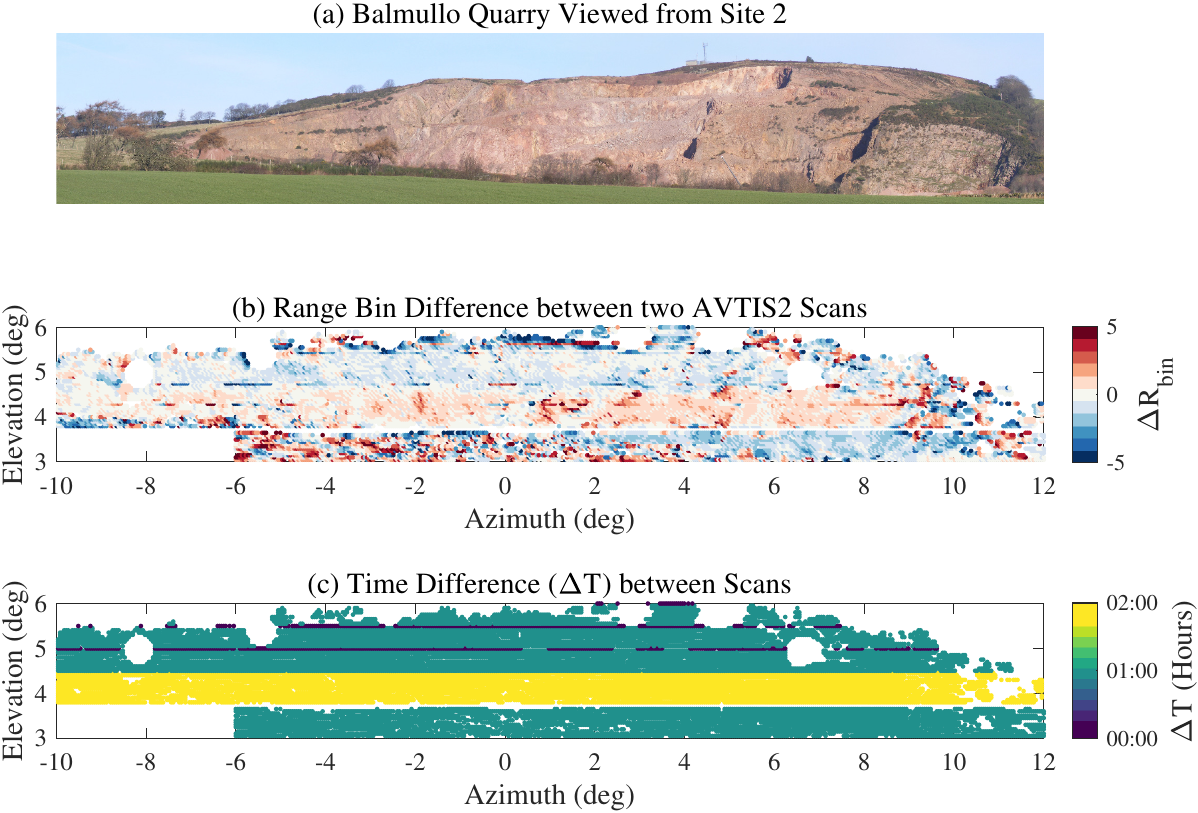}
	\caption{(a) Balmullo Quarry viewed from Site 2, (b) range bin difference $\left(\Delta{R_{\text{bin}}}\right)$ between consecutive scans of the terrain at each angular position across the AVTIS2 Field of View (FoV), and (c) the time difference between measurements at each angular position.}
	\label{AVTIS2_Repeatability_ShortRange}
\end{figure}

% POINT CLOUD REPEATBILITY: SHORT RANGE
A similar spatial pattern in point cloud precision was also observed at Site 2 (1.4 km) (Fig. \ref{AVTIS2_Repeatability_ShortRange}). Here, the range to terrain measured by AVTIS2 varied by $\pm$2 range bins as demonstrated in the histogram of $\Delta{R_{\text{bin}}}$ in Fig. \ref{AVTIS2_Repeatability_Histograms}b. Given the similarity of the $\Delta{R_{\text{bin}}}$ histograms at short- and long-range in Fig. \ref{AVTIS2_Repeatability_Histograms}, the variability in $\Delta{R_{\text{bin}}}$ is independent of range. Where vegetation is present, such as below 4$\degree$ elevation and in the azimuthal range -6$\degree$ to 6$\degree$ (Fig. \ref{AVTIS2_Repeatability_ShortRange}a), $\Delta{R_{\text{bin}}}$ is large and varies between negative and positive values, demonstrating that the range to fluctuating targets such as vegetation will vary significantly between scan acquisitions. Between 3.8$\degree$ and 4.8$\degree$ elevation (Fig. \ref{AVTIS2_Repeatability_ShortRange}), there is a reversal in $\Delta{R_{\text{bin}}}$ from mostly negative to positive values, indicating that the range to terrain is underestimated. This region of terrain was scanned twice $\sim$2 hours apart (Fig. \ref{AVTIS2_Repeatability_LongRange}c), which suggests that the reversal in $\Delta{R_{\text{bin}}}$ was due to a temporal change in the accuracy of AVTIS2 range measurements. There was no precipitation during data acquisition at Site 2 (1.4 km), hence increased atmospheric attenuation is an unlikely cause of the $\Delta{R_{\text{bin}}}$ change. Therefore, the cause of this discrepancy is more likely due to variations in radar hardware altering the AVTIS2 range measurements and subsequently not corrected through range autofocussing.

% POINT CLOUD REPEATBILITY PLOT: SHORT RANGE

% ERROR CONTRIBUTIONS
Point cloud repeatability varies by $\pm$2 range bins at both sites: 1.08 m  at Site 1 (3.3 km) and 1.00 m at Site 2 (1.4 km). These results are independent of georeferencing errors and thus approximate the sensor contribution to AVTIS2 point cloud uncertainty. Taking into account the total point cloud uncertainty using the single point methodology at Site 1 ($\pm$2.75) and Site 2 ($\pm$1.48) from Table \ref{Single_Multiple_Stats}), the relative contribution of sensor errors to point cloud uncertainty can therefore be summarised as follows:
\begin{itemize}
	\item{Site 1 (3.3 km): 39\%}
	\item{Site 2 (1.4 km): 68\%}
\end{itemize}
Whilst this is a crude estimate, it suggests that AVTIS2 sensor characteristics are responsible for over half the total uncertainty at short range but under half at long range. Therefore, long-range point cloud uncertainty is primarily a function of georeferencing uncertainties and terrain properties.

\subsection{Point Cloud Filtering Performance}
% PERFORMANCE
For the single-point processing point cloud, the Voronoi-based outlier detection method is iterated until the number of points removed converges to 0, thus revealing the underlying stable point cloud representing terrain. The iterative routine typically removes visible outliers within the first few iterations, whilst more subtle outliers embedded within the point cloud are removed in subsequent iterations. The algorithm takes $\sim$30-60 seconds for a point cloud with 10,000 points and becomes slower with larger point clouds as the iterative routine takes longer to converge. For the multiple-point processing point cloud, the same iterative routine is applied but with the additional condition that points represented by the spatially filtered single point processing point cloud cannot be removed as these have already been defined as terrain points. An example of a filtered point cloud is shown in Fig. \ref{PointCloud_Filtering_Example}c, illustrating how spatially isolated points are efficiently detected and removed whilst preserving the 3D shape of the terrain. In particular, the new Voronoi-based point cloud filtering algorithm builds upon the workflow described in \cite{Macfarlane2013} for AVTIS2 data and its automation reduces the need for manual filtering of the point cloud data, which is a significant step forward over the prior art. The method is suitable for removing points from sparse point clouds such as those derived from AVTIS2 as it does not require complex plane fitting or calculation of point cloud distance metrics that are computationally intensive. 

\subsection{Georeferencing Accuracy} \label{4_Georeferencing_Performance}
% CC COMBINATIONS
In this section, the effect of changing the number of GCPs in an AVTIS2 scan acquisition on point cloud uncertainty is evaluated. A total of 7 GCPs were used at Site 1 (3.3 km) and 8 GCPs were used at Site 2 (1.4 km) which results in a total of 127 GCP combinations at Site 1 (3.3 km) and 255 GCP combinations at Site 2 (1.4 km).

\begin{figure}[!t]
	\centering
	\includegraphics[width=\linewidth]{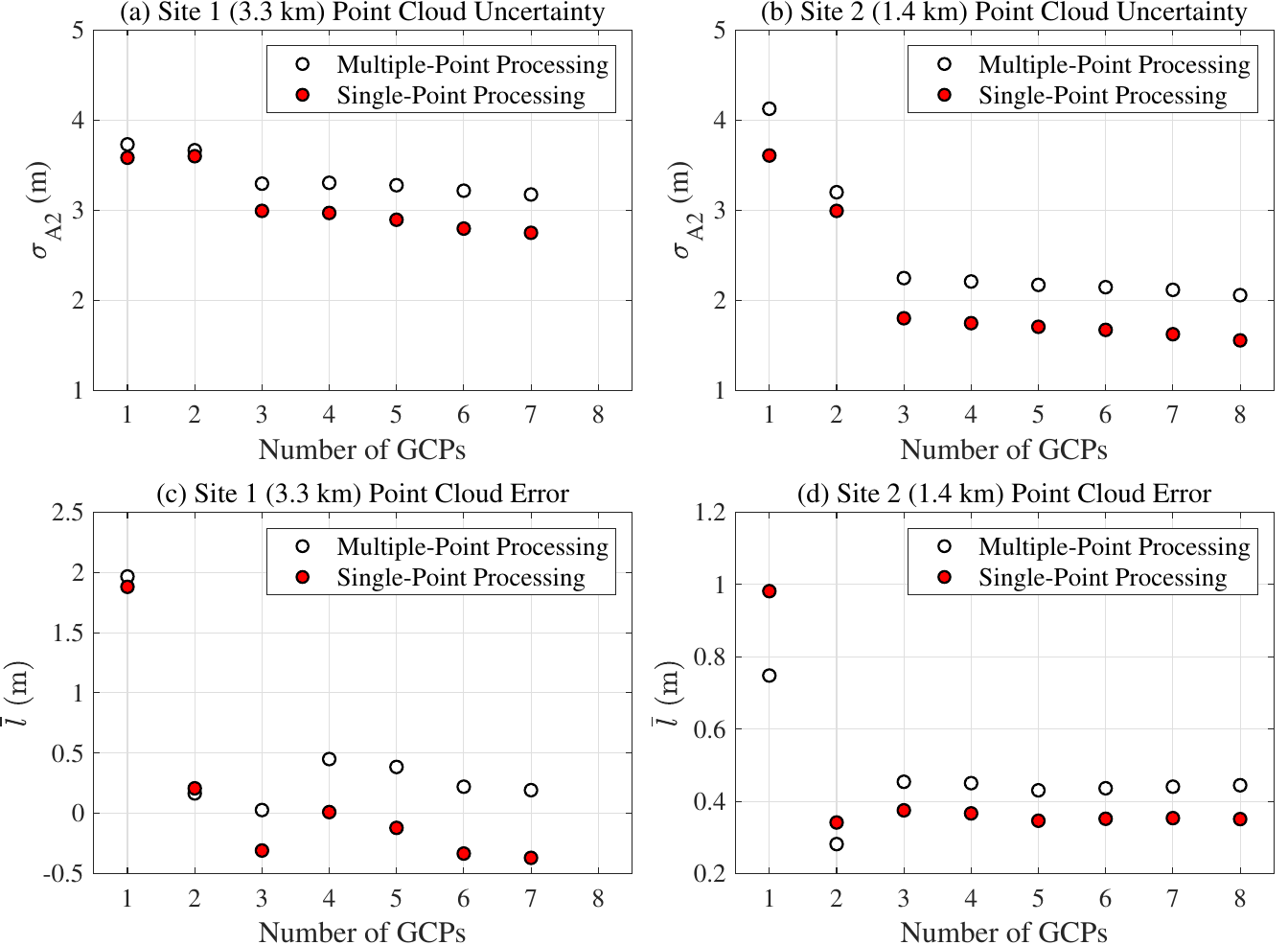}
	\caption{The effect of changing the number of GCPs used for georeferencing AVTIS2 point clouds. The combined uncertainty $\left(\sigma_{\text{A2}}\right)$ is derived by averaging the individual uncertainties from each point cloud, whilst the combined error $(\overline{l})$ is calculated by averaging the individual errors from each point cloud. Uncertainty results are shown for (a) Site 1 (3.3 km) and (b) Site 2 (1.4 km), whilst error results are shown for (c) Site 1 (3.3 km) and (d) Site 2 (1.4 km).}	
	\label{GCP_Number_Performance}
\end{figure}

\begin{figure}[!t]
	\centering
	\includegraphics[width=\linewidth]{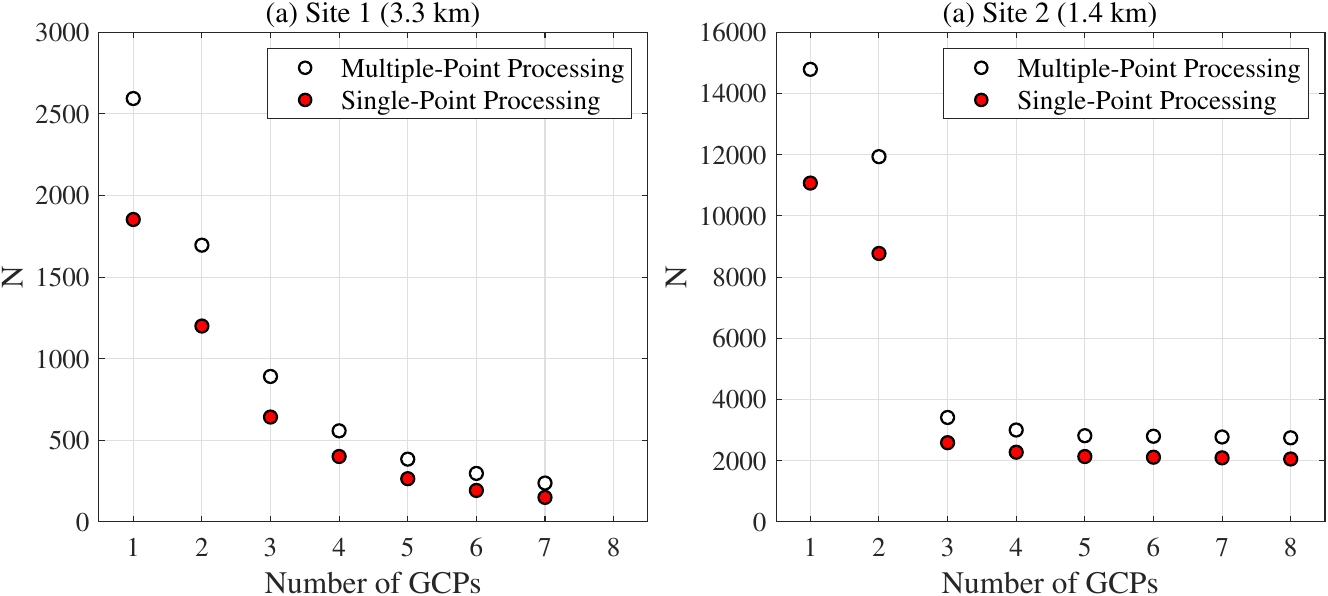}
	\caption{The number of points removed during the M3C2 point cloud comparison process $\left(N\right)$ for (a) Site 1 (3.3 km) and (b) Site 2 (1.4 km). These results are presented in terms of number of GCPs and calculated by averaging the number of points removed for each individual point cloud given its corresponding number of GCPs used for georeferencing.}	
	\label{GCP_Number_NaN}
\end{figure}

% GCP EFFECT ON POINT CLOUD ACCURACY: SHORT-RANGE
At Site 1 (3.3 km), using less than 3 GCPs leads to a point cloud uncertainty of $\pm$3.6 m, whereas using 7 GCPs leads to a point cloud uncertainty of $\pm$2.75 m, hence there is a $\pm$0.85 m (23.61\%) reduction in point cloud uncertainty when increasing the number of GCPs used for georeferencing (Fig. \ref{GCP_Number_Performance}). Using fewer GCPs results in a point cloud that is insufficiently rotated from local to geographic coordinates and cannot be matched with a point from the short-range TLS point cloud during the M3C2 point cloud comparison, hence it is removed. The number of points removed during this process is used as a proxy for the degree of under- or over-rotation in the AVTIS2 point cloud and is plotted in Fig. \ref{GCP_Number_NaN}. The number of points removed at the longer range Site 1 (3.3 km) (Fig. \ref{GCP_Number_NaN}a) reduces negatively exponentially as the number of GCPs used for georeferencing increases and plateaus beyond 5 GCPs. Therefore, AVTIS2 point clouds representing terrain at long range ($>$3 km) should where possible use a minimum of 5 GCPs for accurate georeferencing across the full FoV. This rule is suitable for both the single and multiple point extraction methodologies given that the uncertainty difference between the two is 0.4-0.6 m for all GCP combinations (Table \ref{Single_Multiple_Stats}). Finally, the mean offset of both the single and multiple point clouds varies at Site 1 (3.3 km) when using different combinations of GCPs (Fig. \ref{GCP_Number_Performance}c) although it remains within $\pm$0.5 m of the zero-mean when using 2 or more GCPs and is hence negligible.

% GCP EFFECT ON POINT CLOUD ACCURACY: LONG-RANGE
At Site 2 (1.4 km), using less than 3 GCPs leads to an increase in point cloud uncertainty by $>$1 m, whilst increasing the number of GCPs to more than 3 leads to a marginal $\pm$0.25 m reduction in point cloud uncertainty (Fig. \ref{GCP_Number_Performance}b). The relationship between $\sigma_{\text{A2}}$ and the number of GCPs used for georeferencing follows a negative exponential relationship which plateaus after 3 GCPs. The degree of over- or under-rotation in the AVTIS2 point cloud is again indicated by the number of points removed during the M3C2 process and also follows a negative exponential relationship which plateaus after 3 GCPs (Fig. \ref{GCP_Number_NaN}b). The plateauing of both $\sigma_{\text{A2}}$ and the number of points removed during the M3C2 process at 3 GCPs provides a useful lower limit on the number of GCPs required for accurate georeferencing at short range ($<$1.5 km). Therefore, a smaller number of GCPs are required to accurately georeference short range point clouds compared to those representing terrain at longer ranges. Also, using the multiple point processing methodology leads to point clouds with an uncertainty that is $\pm$0.57 m larger than the single point processing methodology (Fig. \ref{GCP_Number_Performance}b). Despite the weaker performance of the multiple point methodology compared to the single point methodology, its uncertainty is $\pm$0.75 smaller than the single point methodology at long range (Fig.s \ref{GCP_Number_Performance}a and \ref{GCP_Number_Performance}b), which is sufficiently small to warrant its usage at short range. Also, the point cloud error in Fig. \ref{GCP_Number_Performance}d varies around 0.4 m, which is larger but more precise (i.e. less variation) than at Site 1 (3.3 km) indicating that the short range data may underestimate the range to terrain.

\section{Discussion} \label{Section_Discussion}
\subsection{Error Model}
\begin{figure}[!t]
	\centering
	\includegraphics[width=\linewidth]{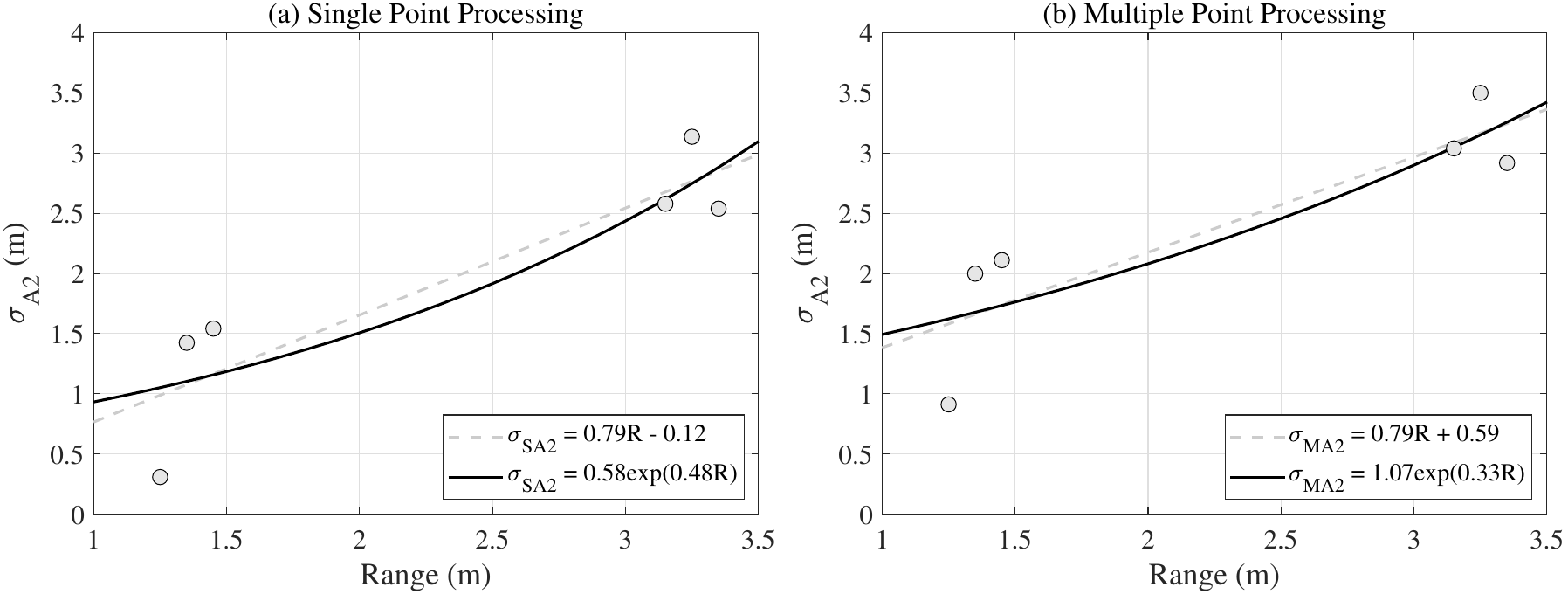}
	\caption{The linear (grey dotted line) and exponential (black line) uncertainty models for the (a) single $\left(\sigma_{\text{SA2}}\right)$ and (b) multiple $\left(\sigma_{\text{MA2}}\right)$ point clouds. The grey dots indicate the standard deviation of the point cloud uncertainty (using all available GCPs) within 100 m range windows. The goodness of fit for both models is given by the adjusted R\textsuperscript{2}$>$0.7 in all four cases.}	
	\label{Linear_Error_Model}
\end{figure}

% RANGE DEPENDENT ERROR MODEL
Several studies have quantified  the accuracy of 3D data sets derived from optical techniques \cite{Lague2013,James2017,Winiwarter2021} and interferometric radar measurements \cite{Noferini2007,Strozzi2012}. This study has now quantified the accuracy of 3D point clouds extracted from 94 GHz radar data cubes for the first time through the development of a new algorithm. The uncertainty of AVTIS2 3D point clouds increases with range when using both the single and multiple point methodologies. By calculating the standard deviation of the point cloud uncertainties in 100 m range windows between 1 km and 3.5 km, a linear regression can be computed between range and $\sigma_{\text{A2}}$ yielding a linear uncertainty model of the form:
\begin{align}
\sigma_{\text{SA2}} = 0.89R - 0.12 \quad \text{ for } \quad 1 \geq R \leq 3.5 \\
\sigma_{\text{MA2}} = 0.79R - 0.59 \quad \text{ for } \quad 1 \geq R \leq 3.5 
\end{align}
where $\sigma_{\text{SA2}}$ is the single point AVTIS2 uncertainty, $\sigma_{\text{MA2}}$ is the multiple point AVTIS2 uncertainty, and $R$ is range in km. The linear uncertainty models for the single and multiple point methodologies are shown are shown in Fig. \ref{Linear_Error_Model} and both have a high degree of correlation (adjusted R\textsuperscript{2}$>$0.7). The model predicts an uncertainty of 0.76 m at 1 km and 2.54 m at 3 km using the single point methodology, whilst it predicts an uncertainty of 1.38 m at 1 km and 3.00 m at 3 km using the multiple point methodology. These predicted uncertainties are in general agreement with the point cloud uncertainties calculated previously, but underestimate point cloud uncertainty at both short and long range. Whilst a linear model fits the available data collected in this study, there are no data points between 1.5 km and 3 km to validate the linearity of the relationship. Wang et al. \cite{Wang2022} found an exponential relationship between uncertainty and range for surface elevation data acquired using the GAMMA Portable Radar Interferometer (GPRI) over Helheim Glacier in Greenland. Therefore, fitting an exponential function through the data in Fig. \ref{Linear_Error_Model} leads to: 
\begin{align}
\sigma_{\text{SA2}} &= 0.58 \exp{\left(0.48R\right)} \quad \text{ for } \quad 1 \geq R \leq 3.5 \\
\sigma_{\text{MA2}} &= 1.07 \exp{\left(0.33R\right)} \quad \text{ for } \quad 1 \geq R \leq 3.5
\end{align}
This exponential model predicts an uncertainty of 0.93 m at 1 km and 2.44 m at 3 km using the single point methodology and an uncertainty of 1.50 m at 1 km and 2.90 m at 3 km using the multiple point methodology. Additional point cloud uncertainty data is required to confirm the exact form of this relationship, but the exponential model is more likely to offer better predictive performance outside the range of values considered here.

% AVTIS2 VS OTHER RADARS
Previous studies have quantified GPRI Digital Elevation Model (DEM) uncertainties of 3.3 m below 2 km \cite{Strozzi2012} and 5 m below 10 km \cite{Wang2022}, both of which are 2 m larger than the short and long range AVTIS2 point cloud uncertainties calculated in this study. Further, the uncertainty of DEMs extracted from Ground Based Synthetic Aperture Radar (GB-SAR) systems can be $\sim$5 m $<$1 km and up to 15 m $>$2 km \cite{Noferini2007,Pieraccini2008} and are thus significantly less accurate than both the GPRI and AVTIS2 3D data sets. Both the GPRI and GB-SAR extract 3D surfaces using phase-based interferometry which is sensitive to changes in atmospheric water vapour. The high sensitivity of both radar systems to atmospheric variability leads to a displacement of the range to terrain in time and is likely the underlying cause of the its larger uncertainty compared to AVTIS2. Therefore, the reduced sensitivity of AVTIS2 to atmospheric variability compared to the phased-based GPRI and GB-SAR instruments supports its application to mapping the 3D geometry of terrain for monitoring purposes.

% AVTIS2 VS TLS
Other studies have reported on the differences between TLS and millimetre-wave radar terrain mapping products. For example, Ryde and Hillier \cite{Ryde2009} measured an accuracy of 0.1-0.3 m below 100 m range for a 95 GHz FMCW radar scanning both a corner cube and a diffuse surface $<$100 m when the environment was both clear and contained obscurants such as dust and rain. In comparison, accuracies of TLS measurements were $<$0.01 m but could not detect targets when mist, rain and dust were present in the atmosphere. Using the same radar system, Hillier et al. \cite{Hillier2015} quantified a 0.2 m difference between millimetre-wave radar and TLS point clouds below 100 m range in a quarry environment. Differences between millimetre-wave radar and TLS are therefore primarily due to the lower range accuracy and larger beam spot size of the radar system compared to TLS.

\subsection{AVTIS2 Point Cloud Uncertainty}
% OVERVIEW OF CONTRIBUTING ERRORS
The uncertainty of AVTIS2 3D point clouds is approximately $\pm$1.5 m below 1.5 km and $\pm$3 m above 3 km.  Where a reference point cloud of a specific study area exists, the AVTIS2 point cloud uncertainty could be reduced by aligning it to the reference data set using techniques such as the Iterative Closest Point (ICP) algorithm \cite{Besl1992}. However, in many circumstances such auxiliary data sets do not exist, hence the results presented in this study should be considered as a benchmark for future studies that use AVTIS2 to map the 3D structure of terrain. The AVTIS2 point cloud uncertainties are a combination of several sources of error that propagate into the final uncertainty estimate:
\begin{enumerate}
	\item{\textbf{Point Cloud Processing}: The point cloud extraction and filtering signal processing methodologies.}
	\item{\textbf{Radar Hardware}: Radar chirp non-linearity, temperature sensitivity and antenna beam pattern.}
	\item{\textbf{Positional Accuracy}: Radar pointing errors, instrument levelling, GCP alignment and radar viewing geometry.}
	\item{\textbf{Environmental Factors}: Atmospheric attenuation, meteorological conditions (e.g. temperature), terrain dielectric properties, and surface roughness.}
\end{enumerate}
The point cloud processing uncertainties were considered extensively in section \ref{Section_Results}, hence the subsequent discussion focusses on points 2-4.

% RADAR HARDWARE
The accuracy of AVTIS2 point clouds is fundamentally limited by the radar hardware. AVTIS2 measurements of range vary over time due to both chirp non-linearity and temperature-dependent range drift \cite{Macfarlane2013,Macfarlane2013a}. These effects are mostly corrected for by using a range autofocussing algorithm that assesses range variability over time by measuring the range to a trihedral reflector whose distance is precisely known using dGPS \cite{Middleton2011}. However, the technique relies upon a physically stable reflector whose centre is pointed directly towards the radar, otherwise its range can be offset from the dGPS distance vector between the reflector and the radar. Also, the radar range to the trihedral reflector should be calculated close in time to a subsequent AVTIS2 scan, otherwise changing atmospheric conditions (e.g. atmospheric water vapour, precipitation, temperature) can lead to a measurement offset that is not incorporated into the range correction. This is particularly significant for temperature-dependent range drift that can systematically alter the AVTIS2 range measurements over time. Therefore, errors resulting from poor range correction may be propagated into the point cloud uncertainty.

% POSITIONAL ACCURACY
For accurate positional measurements, AVTIS2 is levelled on a surveyors' tripod by centring a spirit bubble in order to align the radar line of collimation with the vertical axis. Because the weight of the radar head and gimbal is $\sim$40 kg, and the tripod itself is $\sim$11 kg, the radar set-up is stable against wind buffetting although small radar movement below the angular and range resolution of the radar is possible. However, soft ground and loose footscrews on the AVTIS2 tribrach can lead to gradual changes in tripod levelling over time, which manifest as a gradual change in radar angular measurements over time. Previous studies have found that spirit bubble centering errors have the largest effect at close range (e.g. $<$100 m) \cite{Lambrou2017} and its impact at the ranges of interest in this study (i.e. multiple kilometres) may therefore for be as small as $\sim$1 mm \cite{Lambrou2017} and negligible. Also, radar pointing errors are primarily determined by the stated gimbal accuracy of 0.02$\degree$ which is smaller than the azimuthal and elevation increments of the AVTIS2 radar. Therefore, both tripod levelling and radar pointing errors are expected to be small contributions to georeferencing errors, which is not the case for TLS instruments whose beamwidths are comparatively smaller and hence dynamic changes in instrument position are significant \cite{Hartzell2015}. Instead, point cloud uncertainty is likely to be limited by the radar range resolution $\left(\Delta{R}\right)$. Larger values of $\Delta{R}$ averages terrain over a wider spatial area and will be less accurate than smaller values of $\Delta{R}$. More data is required to confirm both the form and magnitude of this relationship. Finally, the impact of the radar beam pattern is expected to be minimal given that sidelobes are low and the beamwidth is symmetrical in azimuth and elevation.

% ENVIRONMENTAL FACTORS
Both atmospheric and terrain properties impact radar ranging performance and hence point cloud uncertainty. Atmospheric attenuation varies as a result of changes in atmospheric water vapour, obscurants (e.g. dust, fog) and precipitation, altering the radar received power from terrain and hence increase ranging uncertainties. Different surface types also exhibit unique radar backscatter characteristics under different viewing geometries \cite{Ulaby2019} and this also impacts signal stability and hence ranging performance. For example, fluctuating targets such as vegetation are typically considered volume scatterers \cite{Ulaby1990}. As a result, the range to terrain varies within the volumetric medium, hence in the vegetated covered regions of this study point cloud uncertainties were larger. Increased surface roughness may increase the uncertainty of AVTIS2 point clouds. Increased surface roughness reduces the correlation length of a surface \cite{Ulaby2019} and hence may lead to greater variability in radar backscatter which impacts both the magnitude and the shape of the returned echo. This reduces the probability that the range to terrain measurement can be repeated at the same location and hence increases point cloud uncertainty. These influences are largely dependent on the terrain scattering properties, hence future studies should consider point cloud uncertainties over a range of surface types.

\section{Conclusion} \label{Section_Conclusion}
% SUMMARY OF RESULTS
This study has quantified the uncertainty of AVTIS2 3D point clouds using an improved surface extraction methodology based on waveform averaging and developed a new methodology to automatically detect and remove spatial outliers using Voronoi diagrams. The analysis represents the first detailed assessment of millimetre-wave radar point cloud uncertainties and provides a benchmark for interpreting AVTIS2 data sets in future research. The new waveform averaging technique suppresses noise and increases SNR, improving the detectability of terrain returns along a radar range profile. Further, using a signal threshold based on the standard deviation of the averaged signal, multiple terrain surfaces along a range profile can be extracted, which leads to the generation of a more dense point cloud. Total point cloud uncertainties approximately double between 1.5 km and 3 km from $\pm$1.5 m to $\pm$3 m. For a georeferenced point cloud, using a greater number of GCPs reduces point cloud uncertainty. A minimum of 3 GCPs is required for accurate georeferencing at short range ($<$1.5 km) and a minimum of 5 GCPs is required for accurate georeferencing at long range ($>$3 km). The greater number of GCPs required for accurate long range georeferencing is a suggestion, but may not be necessary if a more accurate topographic data set is available for alignment using the Iterative Closest Point (ICP) matching algorithm. Overall, the point clouds and their associated uncertainties quantified in this study are considered sufficient for mapping the 3D geometry of natural terrain, but the uncertainty and precision of the radar must be considered if used for change detection purposes.

% FUTURE WORK
Future work should consider improvements to the AVTIS2 radar and the point cloud extraction process to reduce point cloud uncertainties. Using a Direct Digital Synthesiser (DDS) for chirp generation would overcome the limitations of chirp non-linearity and temperature drift, consequently improving range accuracy and reducing the impact of range drift on radar measurements. Also, the point cloud georeferencing methodology could be improved by employing direct methods which obtain accurate positional information from two displaced GPS antennas at the radar base position. Using DDS and direct georeferencing would reduce the reliance on using GCPs in the field. The development of new surface extraction methods, such as Gaussian Decomposition of AVTIS2 range profiles, may improve the point cloud extraction methodology and hence improve total point cloud accuracy. Further, determining the precise location of the terrain scattering centre within a signal echo is central to the performance of the point cloud processing methodology. Therefore, an improved understanding of the relationship between radar viewing geometry, beam footprint patterns and signal echo characteristics will aid the development of more accurate terrain extraction techniques. Finally, a more robust error model should be developed beyond the ranges of interest considered in this chapter to better constrain the form of the relationship and hence aid in predicting radar performance in future studies.

\section*{Acknowledgments}
William D. Harcourt was funded by the Engineering and Physical Sciences Research Council (EPSRC; grant number: EP/R513337/1) and the Scottish Alliance for Geoscience, Environment and Society (SAGES). The data sets analysed in this paper were collected through a grant awarded by the National Centre for Earth Observation (NCEO) and in collaboration with the University of Glasgow (who collected short-range Terrestrial Laser Scanner (TLS) data), Lancaster University, and the CGG company. The authors are grateful to Eric Murphy, Breedon Aggregates Ltd. for arranging access to the quarry.

%%%%%%%%%%%%%%%%%%%%%%%%%%%%%%%%%%%%%%%%%
\bibliographystyle{IEEEtran}
\bibliography{IEEE_TGARSS_R1_2023_References}
%%%%%%%%%%%%%%%%%%%%%%%%%%%%%%%%%%%%%%%%%

\newpage

\section{Biography Section}
\begin{IEEEbiography}[{\includegraphics[width=1in,height=1.25in,clip,keepaspectratio]{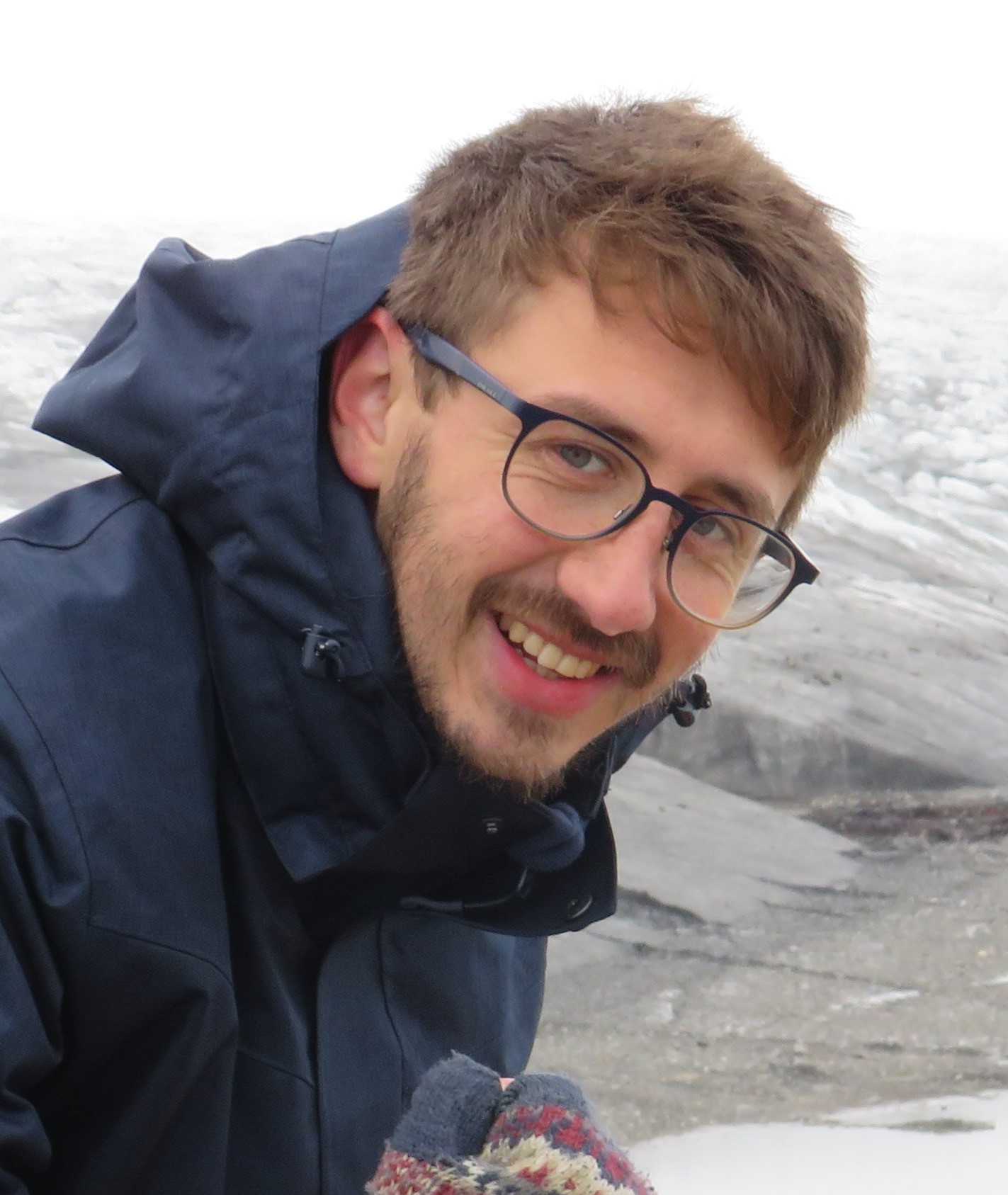}}]{William David Harcourt} received a BSc in Geography from the University of Exeter in 2016 before gaining an MSc in Geographical Information Systems (GIS) from the University of Edinburgh in 2017. He completed his PhD from the University of St Andrews in 2022 with a thesis entitled `The application of millimetre-wave radar to the study of the cryosphere'. His PhD research demonstrated the capabilities of 94 GHz radar for 3D terrain mapping and high resolution monitoring of cryospheric processes such as glacier surface melting, iceberg calving, and snow hazards (e.g. avalanches). Between 2022 and 2023 he was a Lecturer in the School of Geosciences at the University of Aberdeen and from 2023 started a position as an Interdisciplinary Fellow at the University of Aberdeen. His research interests include the development of novel remote sensing methodologies to study ice and snow, the application of machine learning to better understand cryospheric processes, and digital twinning of glaciers and ice sheets.
\end{IEEEbiography}

\vspace{11pt}

\begin{IEEEbiography}[{\includegraphics[width=1in,height=1.25in,clip,keepaspectratio]{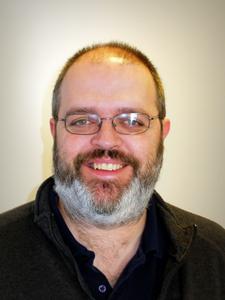}}]{David Graham Macfarlane} received a BSc from the University of Glasgow, Glasgow, UK in 1995, a MSci (Hons) in Theoretical Physics and a PhD in millimeter wave imaging from the University of St Andrews, St Andrews, Fife, UK in 1999 and 2002. Since 2002 he has worked at the Millimetre Wave and EPR Group, University of St Andrews where he is currently a Senior Research Fellow. His research has concentrated on the design, development, build and deployment of a variety of millimetre wave radars for remote sensing, security and laboratory applications. These have included lava dome surface monitoring, real time security radar for people screening and laboratory characterisation of volcanic ash.  He is currently working on the MuWMAS project to measure multi frequency bistatic radar backscatter properties of airborne snow.
\end{IEEEbiography}

\vspace{11pt}

\begin{IEEEbiography}[{\includegraphics[width=1in,height=1.25in,clip,keepaspectratio]{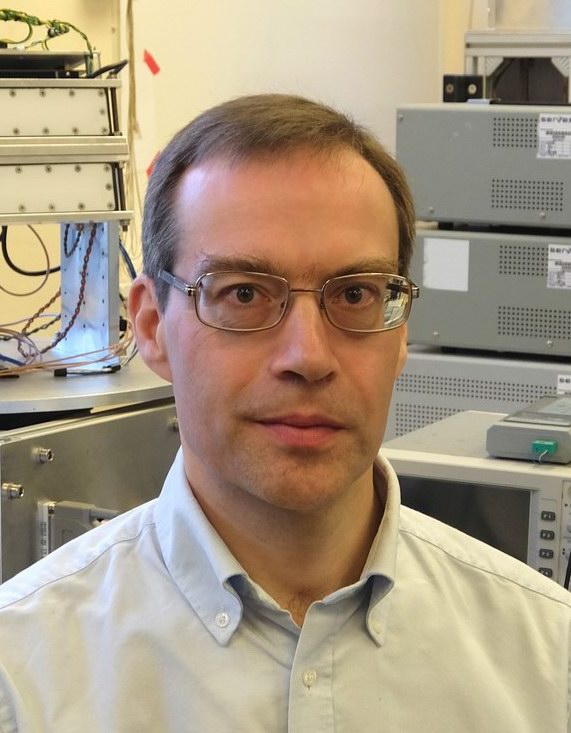}}]{Duncan Alexander Robertson} (Member, IEEE) received the B.Sc. degree (Hons.) in physics and electronics and the Ph.D. degree in millimetre wave physics from the University of St Andrews, U.K., in 1991 and 1994. For most of his career, he has been with the Millimetre Wave Group, University of St Andrews, working on millimetre wave instrumentation. He is currently a Principal Research Fellow in the Group and leads the millimetre wave radar team who develop and field deploy millimeter and sub-millimeter wave radar systems for applications in remote sensing and security. His research interests include high performance coherent radar architectures, low phase noise techniques, target and clutter phenomenology, micro-Doppler signatures for classification, and terrain mapping.
\end{IEEEbiography}

\vspace{11pt}

\vfill

\end{document}